\documentclass[aps,twocolumn,preprintnumbers,amsmath,amssymb,nofootinbib,superscriptaddress,notitlepage]{revtex4}
\usepackage{graphicx,color,dcolumn,booktabs,bm}
\usepackage{longtable,lscape}
\usepackage{txfonts}
\usepackage{overpic}
\usepackage{amssymb}
\usepackage{indentfirst}
\usepackage{feynmf}
\usepackage{slashed}
\usepackage{cases}
\usepackage{multirow}
\usepackage{bbding}
\usepackage{pifont}
\usepackage{appendix}
\usepackage{setspace}
\usepackage{float}
\usepackage[figuresright]{rotating}
\usepackage[colorlinks=true,
            citecolor=blue,
            anchorcolor=red,
            menucolor=red,
            linkcolor=red,
            filecolor=red,
            runcolor=red,
            urlcolor=blue,
            frenchlinks=false]{hyperref}
\usepackage{epstopdf}
\usepackage{siunitx}       
\usepackage{bm}
\usepackage{tabularx}
\usepackage{threeparttable}
\usepackage{bbding}
\usepackage{array}
\usepackage[section]{placeins}
\usepackage{makecell}
\usepackage{comment}
\usepackage{xcolor}
\makeatletter

\newcommand{\Rmnum}[1]{\expandafter\@slowromancap\romannumeral #1@}
\makeatother

\begin{document}
\title{Electromagnetic probes revealing the inner structure of the $\Lambda_c(2940)$}
\author{Ping Chen}
\email{chenp21@lzu.edu.cn}
\affiliation{School of Physical Science and Technology, Lanzhou University, Lanzhou 730000, China}
\affiliation{Research Center for Hadron and CSR Physics, Lanzhou University and Institute of Modern Physics of CAS, Lanzhou 730000, China}
\affiliation{Lanzhou Center for Theoretical Physics, Key Laboratory of Theoretical Physics of Gansu Province, Gansu Provincial Research Center for Basic Disciplines of Quantum Physics, Key Laboratory of Quantum Theory and Applications of MoE, and MoE Frontiers Science Center for Rare Isotopes, Lanzhou University, Lanzhou 730000, China}

\author{Zi-Le Zhang}
\email{zhangzl2023@lzu.edu.cn}
\affiliation{School of Physical Science and Technology, Lanzhou University, Lanzhou 730000, China}
\affiliation{Research Center for Hadron and CSR Physics, Lanzhou University and Institute of Modern Physics of CAS, Lanzhou 730000, China}
\affiliation{Lanzhou Center for Theoretical Physics, Key Laboratory of Theoretical Physics of Gansu Province, Gansu Provincial Research Center for Basic Disciplines of Quantum Physics, Key Laboratory of Quantum Theory and Applications of MoE, and MoE Frontiers Science Center for Rare Isotopes, Lanzhou University, Lanzhou 730000, China}

\author{Yu Zhuge}
\email{220220940231@lzu.edu.cn}
\affiliation{School of Physical Science and Technology, Lanzhou University, Lanzhou 730000, China}
\affiliation{Research Center for Hadron and CSR Physics, Lanzhou University and Institute of Modern Physics of CAS, Lanzhou 730000, China}
\affiliation{Lanzhou Center for Theoretical Physics, Key Laboratory of Theoretical Physics of Gansu Province, Gansu Provincial Research Center for Basic Disciplines of Quantum Physics, Key Laboratory of Quantum Theory and Applications of MoE, and MoE Frontiers Science Center for Rare Isotopes, Lanzhou University, Lanzhou 730000, China}

\begin{abstract}
The $\Lambda_c(2940)$, an open-charm baryon discovered in 2006, has sparked interest due to its ``low mass puzzle'', paralleling the $X(3872)$ in the charmoniumlike sector. Both states challenge conventional hadronic interpretations, with the $X(3872)$ understood as a $D^*\bar{D}$ molecular state and the $\Lambda_c(2940)$ hypothesized as a $D^*N$ molecular state. This work investigates the radiative decay modes $\Lambda_c(2940) \to \Lambda_c(2286)\gamma$, $\Lambda_c(2940) \to \Lambda_c(2595)\gamma$, and $\Lambda_c(2940) \to \Lambda_c(2765)\gamma$, analogous to radiative transitions observed in the $X(3872)$. Using the one-boson-exchange model to obtain the $D^*N$ molecular spatial wave function as input, we calculate decay widths and their ratios, finding differences with different quantum number assumptions. Our findings underscore the potential of electromagnetic probes in revealing its nature and highlight the need for dedicated experimental studies to validate these theoretical predictions.
\end{abstract}

\affiliation{}

\maketitle
\section{Introduction}\label{introduction}
The discovery of the $X(3872)$ \cite{Belle:2003nnu,CDF:2003cab,D0:2004zmu,BaBar:2004iez}, the first identified charmoniumlike $XYZ$ state, has spurred extensive discussions about the novel phenomena associated with the nonperturbative behavior of the strong interaction. This has been further enriched by a series of discoveries of new hadronic states over the past two decades \cite{Swanson:2006st,Chen:2016qju,Chen:2016spr,Esposito:2016noz,Olsen:2017bmm,Brambilla:2019esw,Liu:2019zoy,Chen:2022asf,Liu:2024uxn}. Among these, the $\Lambda_c(2940)$ stands out as a counterpart within the open-charm baryonic sector. It was first observed by the $BABAR$ Collaboration in 2006 through an analysis of the $D^0p$ invariant mass spectrum \cite{BaBar:2006itc} and later confirmed by the Belle Collaboration via its decay channel $\Sigma_c(2455)^{0,++}\pi^{+,-}$ \cite{Belle:2006xni}. The $\Lambda_c(2940)$ shares intriguing similarities with the $X(3872)$. Both exhibit a ``low mass puzzle'', wherein their measured masses are lower than predictions from the quenched potential model if treated as conventional hadrons \cite{Godfrey:1985xj,Yang:2023fsc,Chen:2014nyo,Lu:2018utx,Ebert:2011kk,Capstick:1986ter,Garcilazo:2007eh,Kim:2021ywp}. For the $X(3872)$, this discrepancy has been explained by interpreting it as a $D^*\bar{D}$ molecular state \cite{Tornqvist:1993vu,Swanson:2003tb,Wong:2003xk,Close:2003sg,Voloshin:2003nt,Tornqvist:2004qy,AlFiky:2005jd,Thomas:2008ja,Liu:2008fh,Liu:2009qhy,Lee:2009hy,Braaten:2010mg,Wang:2013kva,Baru:2013rta,Baru:2015nea,Song:2023pdq,Gamermann:2009uq}. Similarly, the $\Lambda_c(2940)$ can be categorized as a $D^*N$ molecular state \cite{Ortega:2012cx,He:2010zq,Zhang:2012jk,Yan:2022nxp,Yan:2023ttx,Xin:2023gkf,He:2006is,Cheng:2006dk,Dong:2009tg,Dong:2010xv,Dong:2011ys,Wang:2015rda,Wang:2020dhf,Dong:2014ksa,Yue:2024paz,Guo:2025efg,Ortega:2013fta,Ortega:2014eoa,Zhang:2019vqe}, with the replacement $\bar{D} \leftrightarrow N$. Notably, the possible $J^P$ quantum numbers for an $S$-wave $D^*N$ system (e.g., $1/2^-$ or $3/2^-$) are richer than those for an $S$-wave $D^*\bar{D}$ system.

The radiative decay modes of the $X(3872)$, such as $X(3872) \to J/\psi\gamma$ and $X(3872) \to \psi(3686)\gamma$, provide critical insights into its inner structure. Experimental measurements of the ratio of these decay widths \cite{BaBar:2008flx,LHCb:2014jvf,Belle:2011wdj,BESIII:2020nbj,LHCb:2024tpv} have been emphasized in recent theoretical studies \cite{Swanson:2004pp,Guo:2014taa,Chen:2024xlw,Barnes:2003vb,Barnes:2005pb,Badalian:2012jz,Li:2009zu,Lahde:2002wj,Mehen:2011ds,Wang:2010ej,Yu:2023nxk,Eichten:2005ga,Dong:2009uf,Cardoso:2014xda}. Drawing on the parallels between the $X(3872)$ and $\Lambda_c(2940)$, it is reasonable to hypothesize that exploring the radiative decay modes of the $\Lambda_c(2940)$ can similarly reveal its internal structure. This serves as the primary motivation for the present study.

Borrowing from the approach used for the $X(3872)$, we identify three potential radiative decay channels for the $\Lambda_c(2940)$: into the ground-state charmed baryon $\Lambda_c(2286)$, its first orbital excitation $\Lambda_c(2595)$, and radial excitation $\Lambda_c(2765)$. This is analogous to the $X(3872)$ decaying into $J/\psi$ and $\psi(3686)$.

To calculate these radiative decay modes, we require the spatial wave function of the $D^*N$ molecular system, a crucial input for this study. Although previous theoretical investigations have addressed this issue \cite{Ortega:2012cx,He:2010zq,Zhang:2012jk,Yan:2022nxp,Yan:2023ttx,Xin:2023gkf,He:2006is,Cheng:2006dk,Dong:2009tg,Dong:2010xv,Dong:2011ys,Wang:2015rda,Wang:2020dhf,Dong:2014ksa}, we perform a comprehensive calculation using the one-boson-exchange (OBE) model to ensure consistency and completeness.

Our analysis focuses on the ratio of the decay widths for $\Lambda_c(2940) \to \Lambda_c(2286)\gamma$, $\Lambda_c(2940) \to \Lambda_c(2595)\gamma$, and $\Lambda_c(2940) \to \Lambda_c(2765)\gamma$. The detailed calculations are presented in subsequent sections. Interestingly, the obtained radiative decay width and ratio vary significantly with possible $J^P$ quantum numbers for an $S$-wave $D^*N$ system.
This discrepancy underscores the utility of electromagnetic probes in unveiling the inner structure of the $\Lambda_c(2940)$. We emphasize that future experimental investigations should prioritize these radiative decay channels to advance our understanding of this state.

This paper is organized as follows. Section \ref{sec2} presents the bound state solutions for the discussed $S$-wave $D^*N$ molecular systems with $J^P=1/2^-$ and $3/2^-$, using the OBE model. In Sec. \ref{sec3}, we examine the radiative decay of the $S$-wave $D^*N$ molecular states. The paper concludes with a brief summary in Sec. \ref{sec5}.

\section{The $S$-wave $D^*N$ molecular states}\label{sec2}
We examine the interactions within the $D^*N$ system mediated by the exchange of light mesons—$\pi$, $\eta$, $\rho$, $\omega$, and $\sigma$—in the OBE model. Following the heavy quark limit and chiral symmetry \cite{Burdman:1992gh, He:2010zq}, the Lagrangians describing the interactions between light pseudoscalar, vector, and scalar mesons with heavy-flavor mesons can be expanded as follows:
\begin{eqnarray}\label{1}
\mathcal{L}_{D^{*}D^{*}\mathbb{P}}& =&
-i\frac{2g}{f_\pi}\varepsilon_{\alpha\mu\nu\lambda}
v^\alpha D^{*\mu}_{b}\partial^\nu \mathbb{P}_{ba}{D}^{*\lambda\dag}_{a},\\[6pt]
\mathcal{L}_{D^{*}D^{*}\mathbb{V}}&=&
-i2\sqrt{2}\lambda{}g_V D^{*\mu}_b (\partial_\mu{}
\mathbb{V}_\nu - \partial_\nu{}\mathbb{V}_\mu)_{ba}D^{*\nu\dag}_a\nonumber\\[6pt]
&~&+\sqrt{2}\beta{}g_V D_b^{*}\cdot D^{*\dag}_a
v\cdot\mathbb{V}_{ba},\\[6pt]
\mathcal{L}_{D^{*}D^{*}\sigma}
&=&2g_sD^{*}_b\cdot{}D^{*\dag}_b\sigma
\end{eqnarray}
with $D^{*T} = (D^{*0}, D^{*+})$. Here, $\mathbb{P}$ and $\mathbb{V}$ represent the pseudoscalar and vector matrices, respectively. The matrix form is given by
\begin{eqnarray}
\mathbb{P}&=&\left(\begin{array}{cc}
{\pi^0 \over \sqrt{2}}+{\eta \over \sqrt{6}} & \pi^+      \\
\pi^-  & -{\pi^0\over \sqrt{2}}+{\eta \over \sqrt{6}} \\
\end{array}
  \right),\\[8pt]
\mathbb{V}^\mu&\!\!=&\left(\begin{array}{cc}
  {\rho^0 \over \sqrt{2}}+{\omega \over \sqrt{2}} & \rho^+      \\
\rho^-  & -{\rho^0\over \sqrt{2}}+{\omega \over \sqrt{2}}   \\
\end{array}\right)^{\mu}.
\end{eqnarray}
The coupling constant $g = 0.59$ is extracted from the total width of $D^{*+}$ \cite{CLEO:2001foe}, and $f_{\pi} = 132$ MeV is the pion decay constant. The couplings $g_v = 5.8$ and $\beta = 0.9$ are determined using the vector meson dominance mechanism \cite{Bando:1987br, Isola:2003fh}, while $\lambda = 0.56 \, \text{GeV}^{-1}$ is obtained by comparing the form factor derived from the light-cone sum rule with the lattice QCD simulation \cite{Casalbuoni:1996pg, Isola:2003fh}. The $\sigma$ meson coupling is given by $g_s = g_{\pi}/(2\sqrt{6})$, where $g_{\pi} = 3.73$ \cite{Liu:2008xz}.

The effective vertices describing the interaction of the nucleon with the pseudoscalar meson $\mathbb{P}$, the vector meson $\mathbb{V}$, and the scalar meson $\sigma$ are
\begin{align}
\mathcal{L}_{NN\mathbb{P}}=&\sqrt{2}g_{\mathbb{P}NN}\bar{N}_bi\gamma_5\mathbb{P}_{ba}N_a,\\[6pt]
\mathcal{L}_{NN\mathbb{V}}=&\sqrt{2}g_{\mathbb{V}NN}\bar{N}_b\gamma_{\mu}\mathbb{V}^{\mu}_{ba}N_a
+\frac{f_{\mathbb{V}NN}}{\sqrt{2}m_N}\bar{N}_b\sigma_{\mu\nu}\partial^\mu\mathbb{V}^\nu_{ba} N_a,\\[6pt]
\mathcal{L}_{NN\sigma}=&g_{\sigma NN}\bar{N}_a\sigma N_a.
\end{align}
Here, $N^T = (n, p)$ represents the nucleon field, and $\sigma_{\mu\nu} = \frac{i}{2} [ \gamma_\mu, \gamma_\nu ]$. The coupling constants are taken from Refs. \cite{Tsushima:1998jz, Engel:1996ic, Machleidt:2000ge, Cao:2010km, He:2010zq} as follows: $g_{\pi NN}^2/(4\pi) = 13.6$, $g_{\eta NN}^2/(4\pi) = 0.4$, $g_{\rho NN}^2/(4\pi) = 0.84$, $g_{\omega NN}^2/(4\pi) = 20$, and $g_{\sigma NN}^2/(4\pi) = 5.69$. The coupling constants $f_{\mathbb{V}NN}$ for the vector mesons $\rho$ and $\omega$ are given by $f_{\rho NN}/g_{\rho NN} = 6.1$ and $f_{\omega NN} = 0$, respectively. In our calculations, the signs of the coupling constants are consistent with the convention in Ref. \cite{Luo:2022cun}. The masses of the mesons and nucleons are taken from the Particle Data Group \cite{ParticleDataGroup:2024cfk}.

Since the $\Lambda_c(2940)$ is an isosinglet, the flavor wave function of the corresponding isoscalar $D^*N$ system with isospin quantum numbers $|I, I_3\rangle$ is given by
\begin{eqnarray}
|\,0,0\rangle=\frac{1}{\sqrt{2}}\left(|D^{*0}p\rangle+|D^{*+}n\rangle\right).
\end{eqnarray}
In the framework of the OBE model, the interaction potentials between the heavy-flavor meson $D^*$ and the nucleon $N$ can be derived through the following steps. Based on the effective Lagrangians that describe the interactions between light mesons, heavy-flavor mesons, and nucleons, the general expressions for the scattering amplitudes of the process $N(p_1)D^*(p_2) \to N(p_3)D^*(p_4)$ via light meson exchange can be written as
\begin{eqnarray}
i\mathcal{M}^E_{ND^*\to ND^*}&=&i\Gamma_{NNE}\,D(q,m_E)\,i\Gamma_{D^*D^*E},
\end{eqnarray}
where $\Gamma_{NNE}$ and $\Gamma_{D^*D^*E}$ represent the interaction vertices for the $NNE$ and $D^*D^*E$ interactions, respectively. The term $D(q, m_E)$ denotes the propagator of the exchanged light mesons. The spinors $u(p_1)$ and $\bar{u}(p_3)$ correspond to the initial and outgoing nucleons, respectively. The Dirac spinor can be written as
\begin{eqnarray}
u(\textbf{p})&=&\sqrt{2m_N}\left(\begin{array}{cc}
  1       \\
\frac{\boldsymbol{\sigma}\cdot\textbf{p}}{2m_N} \\
\end{array}\right)\chi,\\
\bar{u}(\textbf{p})&=&u(\textbf{p})^\dagger\gamma_0=\chi^\dagger\sqrt{2m_N}\left(\begin{array}{cc}
  1&-\frac{\boldsymbol{\sigma}\cdot\textbf{p}}{2m_N} \\[12pt]
\end{array}\right).
\end{eqnarray}
Here, $\chi = (1, 0)^T$ represents spin-up, and $\chi = (0, 1)^T$ represents spin-down.

Next, the effective potentials in momentum space can be related to the obtained scattering amplitude using the Breit approximation \cite{Berestetsky:1982}. The effective potential in momentum space for the $ND^* \to ND^*$ scattering process is
\begin{eqnarray}
  \mathcal{V}_E(\bm{q})&=&-\frac{\mathcal{M}^E_{ND^*\to ND^*}}{\sqrt{ \prod_{i} 2m_i\prod_{f} 2m_f}},
\end{eqnarray}
where $m_i$ and $m_f$ represent the masses of the initial and final states, respectively.

Finally, the effective potentials in coordinate space can be obtained by performing the Fourier transform
\begin{eqnarray}
  \mathcal{V}_E(\bm{r})&=&\int\frac{d^3\bm{q}}{(2\pi)^3}e^{i\bm{q}.\bm{r}}\mathcal{V}_E(\bm{q})\mathcal{F}^2(q^2,m_E^2).
\end{eqnarray}
$\mathcal{F}(q^2, m_E^2) = (\Lambda^2 - m_E^2)/(\Lambda^2 - q^2)$ represents a monopole-type form factor. Here, $\Lambda$ is a phenomenological cutoff parameter, typically on the order of 1 GeV, based on experience with the deuteron \cite{Tornqvist:1993ng, Tornqvist:1993vu}. The explicit $D^*N$ interaction potentials in coordinate space are presented below
\begin{eqnarray}
 V_{\sigma}(r)&=&g_sg_{\sigma NN}\mathcal{A}_{1}Y(\Lambda,m_\sigma,r),\nonumber
 \end{eqnarray}
\begin{widetext}
\begin{eqnarray}
  V_{\mathbb{P}}(r)&=&-\frac{gg_{\pi NN}}{3\sqrt{2}f_\pi m_N}\left[\mathcal{A}_{2}\nabla^2+S(r,i\bm\epsilon_2\times{\bm\epsilon^{\dagger}_4},\sigma)\hat{\mathcal{T}} \right]Y(\Lambda,m_\mathbb{P},r),\nonumber\\
  V_{\mathbb{V}}(r)&=&-\beta g_v g_{\mathbb{V}NN}\mathcal{A}_{1}Y(\Lambda,m_\mathbb{V},r)
  +\frac{g_{\mathbb{V}NN}\lambda g_v}{3m_N}\left[2\mathcal{A}_{2}\nabla^2-S(r,i\bm\epsilon_2\times{\bm\epsilon^{\dagger}_4},\sigma)\hat{\mathcal{T}}\right]Y(\Lambda,m_\mathbb{V},r)\nonumber\\
  &&+\frac{f_{\mathbb{V}NN}\lambda g_v}{3m_N}\left[2\mathcal{A}_{2}\nabla^2-S(r,i\bm\epsilon_2\times{\bm\epsilon^{\dagger}_4},\sigma)\hat{\mathcal{T}}\right]Y(\Lambda,m_\mathbb{V},r)
  -\frac{f_{\mathbb{V}NN}\beta g_v}{4m_N^2}\left[\mathcal{A}_{1}\nabla^2+2\mathcal{A}_{1}\sigma\cdot\bm{L}\hat{Q}\right]Y(\Lambda,m_\mathbb{V},r).
\end{eqnarray}
\end{widetext}

\setlength{\tabcolsep}{2.2mm}
\begin{table}[tbp]
\caption{Bound state solutions for the isoscalar $D^*N$ systems with quantum numbers $J^P = 1/2^-$ and $3/2^-$ are presented. Here, $\Lambda$ is the cutoff parameter, and $E$ denotes the binding energy. $M_{\rm mol}$ and $r_{\rm RMS}$ refer to the mass and root-mean-square radius of the bound state $D^*N$, respectively.}\label{binding}
\renewcommand\arraystretch{1.35}
\centering
\setlength{\tabcolsep}{0.75mm}
\begin{tabular*}{85mm}{@{\extracolsep{\fill}}cc
ccc}

\toprule[1pt]
\toprule[1pt]
$J^{P}$&$\Lambda\,($GeV$)$&\textit{E}\,(MeV)&$M_{\rm mol}$\,(MeV)&$r_{\rm RMS}$\,(fm)\\

\toprule[0.75pt]
\multirow{3}{*}{$\frac{1}{2}^{-}$}
&1.35&$-1.36$&2946.11&3.73\\
~&1.37&$-5.51$&2941.96&1.95\\
~&1.40&$-18.98$&2928.49&1.11\\
\toprule[0.75pt]
\multirow{3}{*}{$\frac{3}{2}^{-}$}
&1.08&$-1.45$&2946.02&3.55\\
~&1.17&$-6.06$&2941.41&1.94\\
~&1.37&$-20.70$&2926.77&1.25\\

\bottomrule[1pt]
\bottomrule[1pt]
\end{tabular*}
\end{table}

Here, the shorthand notations for the operators are $\mathcal{A}_{1}=\bm\epsilon_2\cdot{\bm\epsilon^{\dagger}_4}$ and $\mathcal{A}_{2}=i\bm\epsilon_2\times{\bm\epsilon^{\dagger}_4}\cdot\bm\sigma$. The tensor operator is defined as $S(\bm{r},{\bf A},{\bf B})=3{\bf A}\cdot\hat{r}{\bf B}\cdot\hat{r}-{\bf A}\cdot{\bf B}$. The corresponding matrix elements $\langle f | \mathcal{A}_k | i \rangle$ can be obtained by sandwiching these operators between the initial and final spin-orbit wave functions. For the $D^*N$ system discussed, the corresponding spin-orbit wave function is given by
\begin{eqnarray}\label{spin-orbit}
D^*N:|\,{}^{2S+1}L_{J}\rangle\sim\!\!\!\!\sum\limits_{m,m^\prime,m_S,m_L}\!\!\!\!C_{\frac{1}{2}m,1m^\prime}^{S,m_S}C^{J,M}_{Sm_S,L m_L}\chi_{\frac{1}{2}m}\epsilon^{m^\prime}\,Y_{L,m_L},
\end{eqnarray}
where $C_{\frac{1}{2}m,1m^\prime}^{S,m_S}$ and $C^{J,M}_{Sm_S,L m_L}$ are the Clebsch-Gordan coefficients, $Y_{L,m_L}$ is spherical harmonics function.
The function $Y(\Lambda,m_i,r)$ is written as
\begin{eqnarray}
Y(\Lambda,m_i,r)&=&\frac{e^{-m_i r}-e^{-\Lambda r}}{4\pi r}-\frac{\Lambda^2-m^2_i}{8\pi\Lambda}e^{-\Lambda r},
\end{eqnarray}
and operators $\nabla^2$, $\hat{\mathcal{T}}$ and $\hat{Q}$ acting on $Y(\Lambda,m_i,r)$ have the forms
\begin{eqnarray}
\nabla^2Y(\Lambda,m_i,r)&=&\frac{1}{r^2}\frac{\partial}{\partial r}r^2\frac{\partial}{\partial r}Y(\Lambda,m_i,r),\\[6pt]
\hat{\mathcal{T}}Y(\Lambda,m_i,r)&=&r\frac{\partial}{\partial r}\frac{1}{r}\frac{\partial}{\partial r}Y(\Lambda,m_i,r),\\[6pt]
\hat{Q}Y(\Lambda,m_i,r)&=&\frac{1}{r}\frac{\partial}{\partial r}Y(\Lambda,m_i,r).
\end{eqnarray}
With the preparations outlined above, the total interaction potential for the $D^*N$ system with isospin $I = 0$ can be expressed as
\begin{eqnarray}
   V(r)&=&V_{\sigma}(r)+\mathcal{G}\left(V_{\pi}(r)+V_{\rho}(r)\right)+\mathcal{H}\left(V_{\omega}(r)+\frac{1}{3}V_{\eta}(r)\right),\nonumber
\end{eqnarray}
where the isospin factors are $\mathcal{G} = 3/2$ and $\mathcal{H} = 1/2$.

Once the interaction potentials in coordinate space for the isoscalar $D^*N$ system are obtained, bound state solutions, including the binding energy $E$ and the root-mean-square radius $r_{\rm RMS}$, can be obtained by solving the Sch\"odinger equation and with the Gaussian expansion method \cite{Hiyama:2003cu, Hiyama:2012}.

Next, we present the bound state solutions for the isoscalar $D^*N$ system with $J^P = 1/2^-$ and $3/2^-$, denoted as $[D^*N]_{1/2^-}$ and $[D^*N]_{3/2^-}$ for simplicity, respectively. The bound state solutions for the $[D^*N]_{1/2^-}$ and $[D^*N]_{3/2^-}$ states are presented in Table \ref{binding}. For the $[D^*N]_{1/2^-}$ state, a bound state is obtained with a binding energy of $-5.51$ MeV and a root-mean-square radius of 1.95 fm for a cutoff parameter of 1.37 GeV. At this cutoff, the mass of the bound state is close to that of the $\Lambda_c(2940)$. For the $[D^*N]_{3/2^-}$ state, the bound state has a binding energy of $-6.06$ MeV and a root-mean-square radius of 1.94 fm at a cutoff of 1.17 GeV. The corresponding mass $M_{\rm mol}$ is near the $\Lambda_c(2940)$.\footnote{The LHCb measurement indicates that the $\Lambda_c(2940)$ is favored to have $J^P = 3/2^-$ \cite{LHCb:2017jym}.}

\begin{figure}[tbp]
\includegraphics[width=245pt]{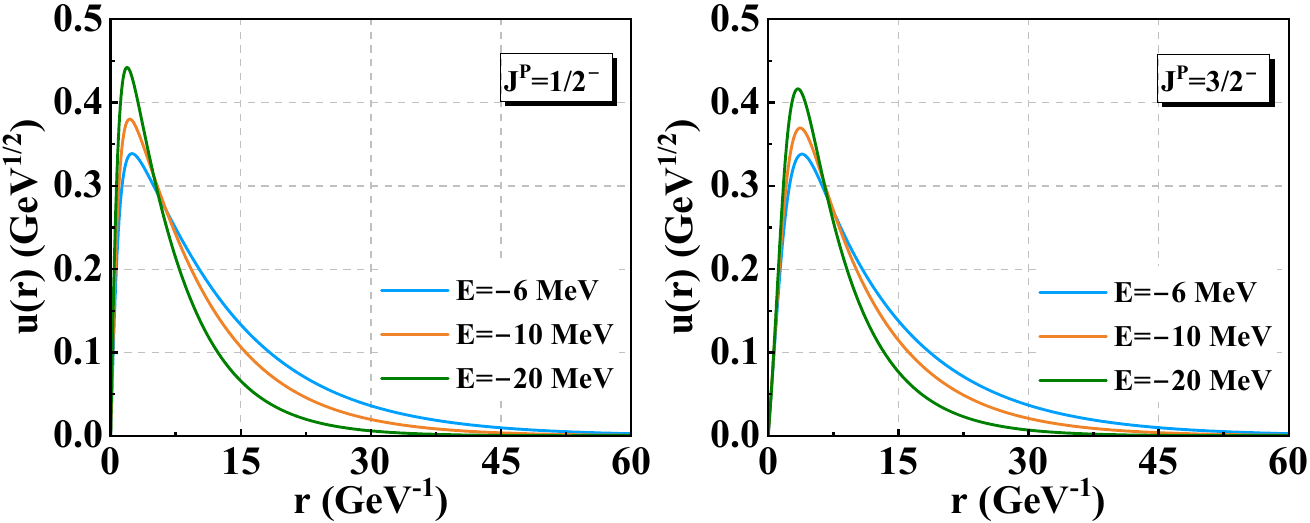}
\caption{Radial component of the spatial wave function of $S$-wave $D^*N$ molecular states with $J^P=1/2^-$ and $3/2^-$ in coordinate space. Here, we take several typical values of binding energy, $-6$ MeV, $-10$ MeV, and $-20$ MeV. }\label{wavefunction}
\end{figure}

The radial component of the obtained spatial wave functions of the bound states $[D^*N]_{1/2^-}$ and $[D^*N]_{3/2^-}$, with typical binding energies of $-6$ MeV, $-10$ MeV, and $-20$ MeV, are shown in Fig. \ref{wavefunction}. These results will be used as input for calculating the radiative decays discussed in the following section.

\section{Three typical radiative decay modes of the $S$-wave $D^*N$ molecular states}\label{sec3}

In this section, we focus on the radiative decay of a potential $D^*N$ molecular state. Assuming the $J^P$ quantum numbers of the $D^*N$ system are $1/2^-$ and $3/2^-$, we calculate the radiative decay processes for the transitions $D^*N \to \Lambda_c(2286) \gamma$, $D^*N \to \Lambda_c(2595) \gamma$, and $D^*N \to \Lambda_c(2765) \gamma$ using the effective Lagrangian approach.

\begin{figure*}[tbp]
\includegraphics[width=500pt]{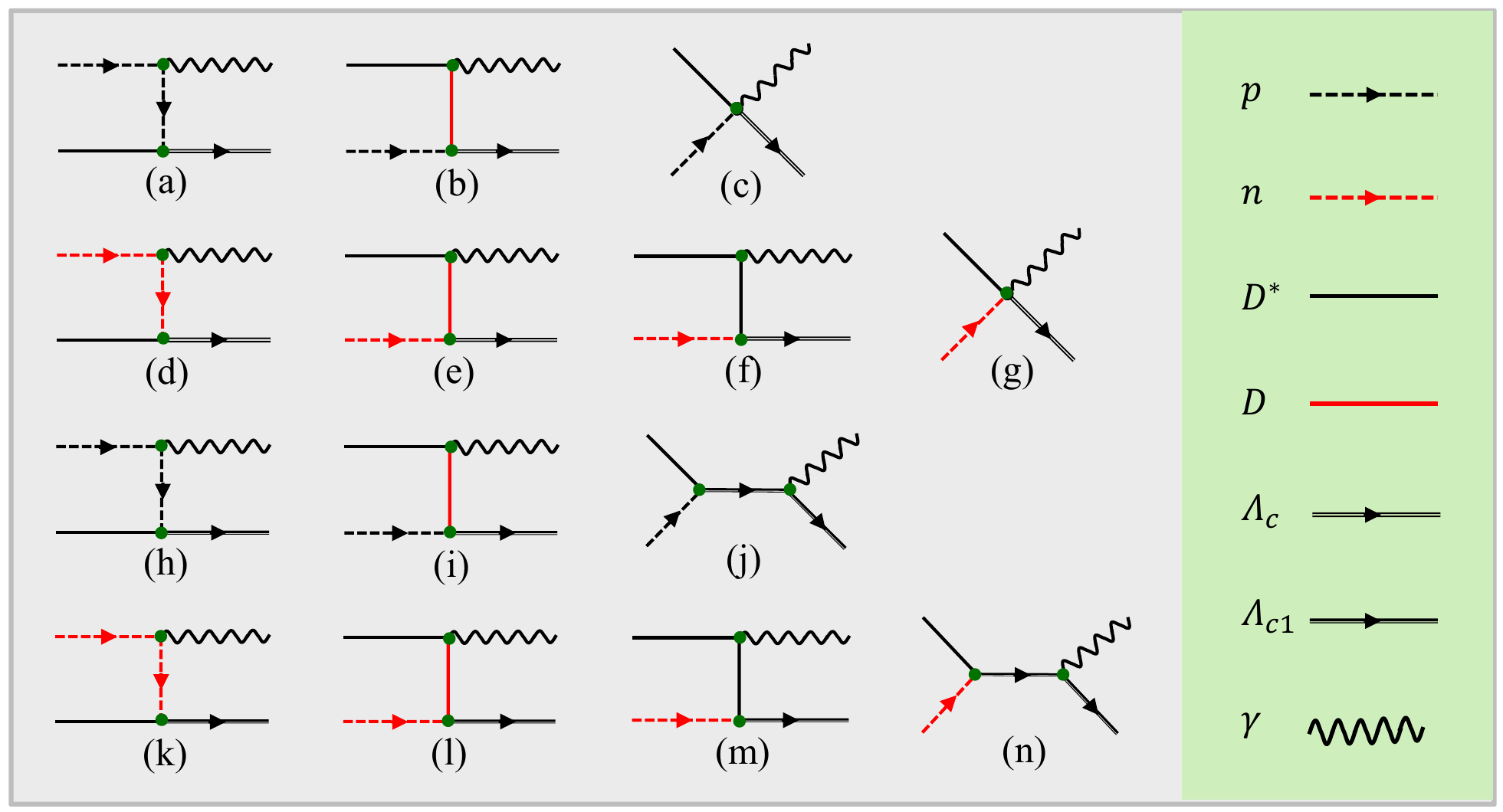}
\caption{(Color online.) Diagrams illustrating the transition from the $D^*N$ molecular state to the final states $\Lambda_c(2286) \gamma$ and $\Lambda_c(2595) \gamma$.}\label{diagram}
\end{figure*}

The Feynman diagrams representing the processes $D^*N \to \Lambda_c(2286) \gamma$, $D^*N \to \Lambda_c(2595) \gamma$ and $D^*N \to \Lambda_c(2765) \gamma$ are shown in Fig. \ref{diagram}, which includes the vertices for both strong and electromagnetic interactions. For brevity, we will refer to $\Lambda_c(2286)$, $\Lambda_c(2595)$, and $\Lambda_c(2765)$ as $\Lambda_c$, $\Lambda_{c1}$, and $\Lambda_{c2}$, respectively, in the following discussion.\footnote{The $\Lambda_c(2286)$ is a $1S$ state with $J^P=1/2^+$, while the $\Lambda_c(2595)$ corresponds to a $1P$ state with $J^P=1/2^-$, and the $\Lambda_c(2765)$ corresponds to a $2S$ state with $J^P=1/2^+$ \cite{Ebert:2011kk,Chen:2014nyo,Chen:2016iyi,Wang:2020mxk,Zhang:2024afw,Lu:2016ctt}.} Notably, the $\Lambda_c(2595)$ and
 $\Lambda_c(2765)$ represent the first orbital and the first radial excitation of the $\Lambda_c(2286)$, respectively.

The effective Lagrangians corresponding to the vertices of $ND^{(*)}\Lambda_c$ and $ND^{(*)}\Lambda_{c(1,2)}$ are \cite{Dong:2010xv,Dong:2009tg,He:2011jp,Dong:2014ksa}
\begin{eqnarray}\label{strong1}
\hspace*{-2cm}
{\cal L}_{ND\Lambda_{c(2)}}&=&ig_{ND\Lambda_{c(2)}}\bar{\Lambda}_{c(2)}\gamma_5ND+ \rm H.c.,\\[6pt]
{\cal L}_{ND^*\Lambda_{c(2)}}&=&g_{ND^*\Lambda_{c(2)}}\bar{\Lambda}_{c(2)}\gamma^\mu ND^*_\mu+ \rm H.c.,\\[6pt]
{\cal L}_{ND\Lambda_{c1}}&=&ig_{ND\Lambda_{c1}}\bar{\Lambda}_{c1} ND+\rm H.c.,\\[6pt]
{\cal L}_{ND^*\Lambda_{c1}}&=&g_{ND^*\Lambda_{c1}}\bar{\Lambda}_{c1}\gamma^\mu\gamma^5ND^*_\mu+ \rm H.c.,\label{strong4}
\end{eqnarray}

We use the quark-pair-creation model to determine the above coupling constants in the Appendix \cite{Qiao:2024acm,Zhang:2022pxc}. The results are shown in Table \ref{coupling}. They are more or less consistent with the results given by the SU(4) symmetry ($g_{ND\Lambda_c}=13.5$ and $g_{ND^*\Lambda_c}=-5.6$ \cite{Lin:1999ve,Liu:2001ce,Dong:2009tg}) and the QCD sum rule ($g_{ND\Lambda_c}=7.9$ and $g_{ND^*\Lambda_c}=-7.5$ \cite{Khodjamirian:2011jp,Navarra:1998vi,Duraes:2000js}).
\setlength{\tabcolsep}{2.2mm}
\begin{table}[htbp]
\caption{Values of coupling constants involved in the interaction of $D^{(*)} N \Lambda_{c(1,2)}$.}\label{coupling}
\renewcommand\arraystretch{1.45}
\centering
\setlength{\tabcolsep}{0.75mm}
\begin{tabular*}{85mm}{@{\extracolsep{\fill}}cc
ccccc}
\toprule[1pt]
\toprule[1pt]
Couplings&$g_{DN\Lambda_c}$&$g_{D^*N\Lambda_c}$&$g_{DN\Lambda_{c1}}$&$g_{D^*N\Lambda_{c1}}$&$g_{DN\Lambda_{c2}}$&$g_{D^*N\Lambda_{c2}}$\\
\toprule[0.75pt]
Values&8.5&$-6.0$&$5.9$&$-$2.0&$4.3$&$-2.4$\\
\bottomrule[1pt]
\bottomrule[1pt]
\end{tabular*}
\end{table}

The electromagnetic component of the Lagrangian includes the following terms \cite{Dong:2010xv,Dong:2009uf}:
\begin{align}
{\cal L}_{\Lambda_{c1}\Lambda_{c1}\gamma}=&\,\,e\bar{\Lambda}_{c1}A_\mu\gamma^\mu\Lambda_{c1},\\[6pt]
{\cal L}_{NN\gamma}=&\,\,e\bar{N}\left(A_\mu\gamma^\mu Q_N+F_{\mu\nu}\sigma^{\mu\nu}\frac{\kappa_N}{4m_N}\right)N,\\[6pt]
{\cal L}_{D^*D^*\gamma}=&-ieA_\mu\bigg(g^{\alpha\beta}\!
D^{*-}_\alpha \!\stackrel{\leftrightarrow}{\partial^{\,\mu}}\!
 D^{*+}_\beta  \nonumber\\[6pt]
&-g^{\mu\beta} D^{*-}_\alpha \partial^\alpha D^{*+}_\beta
+ g^{\mu\alpha}\partial^\beta D^{*-}_\alpha D^{*+}_\beta\bigg) ,\\[6pt]
{\cal L}_{D^*D\gamma}=&\,\,\frac{e}{4}\epsilon^{\mu\nu\alpha\beta}F_{\mu\nu}\left(g_{D^{*-}D^+\gamma} D^{*-}_{\alpha\beta}D^++g_{D^{*0}D^0\gamma}\bar{D}^{*0}_{\alpha\beta}D^0\right)\nonumber\\
&+\rm H.c..
\end{align}
Here, $Q_N$ represents the nucleon charge, with $Q_p = 1$ for the proton and $Q_n = 0$ for the neutron. The anomalous magnetic moments of the proton and neutron are $\kappa_p = 1.793$ and $\kappa_n = -1.913$, respectively. The photon field is denoted by $A_\mu$, and the electromagnetic field tensor is given by $F_{\mu\nu} = \partial_\mu A_\nu - \partial_\nu A_\mu$. The term $D^*_{\alpha\beta}$ is defined as $\partial_\alpha D^*_\beta-\partial_\beta D^*_\alpha$, while $D^*_\alpha\stackrel{\leftrightarrow}{\partial}_{\mu}\! D^*_\beta = D^*_\alpha \partial_{\mu} D^*_\beta - \partial_{\mu} D^*_\alpha D^*_\beta$. The coupling constant $g_{D^*D\gamma}$ is determined using experimental data on the radiative decay widths $\Gamma(D^{*} \to D \gamma)$ \cite{Dong:2010xv}, which are
\begin{eqnarray}
g_{D^{*+}D^+\gamma}=0.5~{\rm{GeV^{-1}}},\,\,g_{D^{*0}D^0\gamma}=2.0~\rm{GeV^{-1}}.\nonumber
\end{eqnarray}
According to the Feynman rules, the scattering amplitude for the process $N(p_1)D^*(p_2) \to \gamma(k_1)\Lambda_c(k_2)$ can be expressed as
\begin{align}\label{Ma}
i\mathcal{M}^{(a)}=&-e\left[g_{ND^*\Lambda_c}\bar{u}(k_2)\gamma_\alpha\epsilon^{\alpha}_{D^*}\right]\frac{i(q \mkern -9.5 mu/_1+m_p)}{q_1^2-m_p^2}\nonumber\\
&\times\left[\epsilon^{*\beta}_\gamma\gamma_\beta+i(k_{1}^\mu\epsilon^{*\nu}_\gamma-k_{1}^\nu\epsilon^{*\mu}_\gamma)\,\sigma_{\mu\nu}\frac{\kappa_p}{4m_N}\right]u(p_1),\\[12pt]
i\mathcal{M}^{(b)}=&-ieg_{D^{*0}D^0\gamma}g_{ND\Lambda_c}\bar{u}(k_2)\gamma_5u(p_1)\varepsilon_{\mu\lambda\alpha\rho}k_{1}^\mu\epsilon^{*\lambda}_\gamma p_{2}^\alpha\epsilon^\rho_{D^*}\nonumber\\
&\times\frac{i}{q_2^2-m_{D^0}^2}\label{ amplitude1},
\end{align}
\begin{align}\label{Mc}
i\mathcal{M}^{(c)}=&\,\,ieg_{ND^*\Lambda_c}\bar{u}(k_2)\gamma_\mu\epsilon^\mu_{D^*}\frac{(2p_1-k_1)_\nu}{q_1^2-m_{p}^2}\epsilon^{*\nu}_\gamma u(p_1),
\end{align}
\begin{align}
i\mathcal{M}^{(d)}=&-ieg_{ND^*\Lambda_c}\bar{u}(k_2)\gamma_\alpha\epsilon^\alpha_{D^*}\frac{i(q \mkern -9.5 mu/_1+m_n)}{q_1^2-m_n^2}(k_{1}^\mu\epsilon^{*\nu}_\gamma-k_{1}^\nu\epsilon^{*\mu}_\gamma)\nonumber\\
&\times\sigma_{\mu\nu}\frac{\kappa_n}{4m_N}u(p_1),
\end{align}
\begin{align}
i\mathcal{M}^{(e)}=&-ieg_{D^{*+}D^+\gamma}g_{ND\Lambda_c}\bar{u}(k_2)\gamma_5u(p_1)\varepsilon_{\mu\lambda\alpha\rho}k_{1}^\mu\epsilon^{*\lambda}_\gamma p_{2}^\alpha\epsilon^\rho_{D^*}\nonumber\\
&\times\,\frac{i}{q_2^2-m_{D^+}^2},
\end{align}
\begin{align}
i\mathcal{M}^{(f)}=&-eg_{ND^*\Lambda_c}\bar{u}(k_2)\gamma_\rho{u}(p_1)\epsilon^{*\lambda}_\gamma\nonumber\\
&\times\left[(-q_2-p_2)_\lambda g_{\gamma\sigma}+p_{2\sigma}g_{\gamma\lambda}\right.\nonumber\\
&\left.+q_{2\gamma} g_{\lambda\sigma}\right]\epsilon_{D^*}^\gamma\frac{i(-g^{\rho\sigma}+q_{2}^\rho q_{2}^\sigma/m_{D^*}^2)}{q_2^2-m_{D^*}^2},
\end{align}
\begin{align}\label{Mg}
i\mathcal{M}^{(g)}=&\,\,ieg_{ND^*\Lambda_c}\bar{u}(k_2)\gamma_\mu\epsilon^\mu_{D^*}\frac{(2p_2-k_1)_\nu}{q_2^2-m_{D^*}^2}\epsilon^{*\nu}_\gamma u(p_1),
\end{align}
where Eq. (\ref{Mc}) and Eq. (\ref{Mg}) are contact terms, and the generalized contact
term is introduced in Refs. \cite{Haberzettl:2006bn,Wang:2015rda,Huang:2016tcr,Huang:2011as,Gao:2010hy}, which maintains the gauge invariance of the photon field. The scattering amplitudes of $N(p_1)D^*(p_2)\to\gamma(k_1)\Lambda_{c2}(k_2)$ process can be obtained by replacing the coupling constant $g_{D^{(*)}N\Lambda_c}$ with $g_{D^{(*)}N\Lambda_{c2}}$ in Eqs. (\ref{Ma}$)-($\ref{Mg}).

For the process of $N(p_1)D^*(p_2)\to\gamma(k_1)\Lambda_{c1}(k_2)$, the scattering amplitudes can be written as
\begin{align}\label{Maa}
i\mathcal{M}^{(h)}=&-e\left[g_{ND^*\Lambda_{c1}}\bar{u}(k_2)\gamma_\alpha\gamma_5\epsilon^{\alpha}_{D^*}\right]\frac{i(q \mkern -9.5 mu/_1+m_p)}{q_1^2-m_p^2}\nonumber\\
&\times\left[\epsilon^{*\beta}_\gamma\gamma_\beta+i(k_{1}^\mu\epsilon^{*\nu}_\gamma-k_{1}^\nu\epsilon^{*\mu}_\gamma)\,\sigma_{\mu\nu}\frac{\kappa_p}{4m_N}\right]u(p_1),\\[12pt]
i\mathcal{M}^{(i)}=&-ieg_{D^{*0}D^0\gamma}g_{ND\Lambda_{c1}}\bar{u}(k_2)u(p_1)\varepsilon_{\mu\lambda\alpha\rho}k_{1}^\mu\epsilon^{*\lambda}_\gamma p_{2}^\alpha\epsilon^\rho_{D^*}\nonumber\\
&\times\frac{i}{q_2^2-m_{D^0}^2},
\end{align}
\begin{align}
i\mathcal{M}^{(j)}=&-eg_{ND^*\Lambda_{c1}}\bar{u}(k_2)\epsilon^{*\lambda}_\gamma\gamma_\lambda\frac{i(q \mkern -9.5 mu/_3+m_{\Lambda_c})}{q_3^2-m_{\Lambda_c}^2}\gamma_\mu\gamma_5\epsilon^\mu_{D^*} u(p_1),
\end{align}
\begin{align}
i\mathcal{M}^{(k)}=&-ieg_{ND^*\Lambda_{c1}}\bar{u}(k_2)\gamma_\alpha\gamma_5\epsilon^\alpha_{D^*}\frac{i(q \mkern -9.5 mu/_1+m_n)}{q_1^2-m_n^2}(k_{1}^\mu\epsilon^{*\nu}_\gamma-k_{1}^\nu\epsilon^{*\mu}_\gamma)\nonumber\\
&\times \sigma_{\mu\nu}\frac{\kappa_n}{4m_N}u(p_1),\\[12pt]
i\mathcal{M}^{(l)}=&-ieg_{D^{*+}D^{+}\gamma}g_{ND\Lambda_{c1}}\bar{u}(k_2)u(p_1)\varepsilon_{\mu\lambda\alpha\rho}k_{1}^\mu\epsilon^{*\lambda}_\gamma p_{2}^\alpha\epsilon^\rho_{D^*}\nonumber\\
&\times\,\frac{i}{q_2^2-m_{D^+}^2},
\end{align}
\begin{align}
i\mathcal{M}^{(m)}=&-eg_{ND^*\Lambda_{c1}}\bar{u}(k_2)\gamma_\rho\gamma_5{u}(p_1)\epsilon^{*\lambda}_\gamma\nonumber\\[5pt]
&\times\left[(-q_2-p_2)_\lambda g_{\gamma\sigma}+p_{2\sigma}g_{\gamma\lambda}+q_{2\gamma} g_{\lambda\sigma}\right]\nonumber\\[5pt]
&\times\epsilon^\gamma_{D^*}\frac{i(-g^{\rho\sigma}+q_{2}^\rho q_{2}^\sigma/m_{D^*}^2)}{q_2^2-m_{D^*}^2},\\[12pt]
i\mathcal{M}^{(n)}=&-eg_{ND^*\Lambda_{c1}}\bar{u}(k_2)\epsilon^{*\lambda}_\gamma\gamma_\lambda\frac{i(q \mkern -9.5 mu/_3+m_{\Lambda_c})}{q_3^2-m_{\Lambda_c}^2}\gamma_\mu\gamma_5\epsilon^\mu_{D^*}u(p_1)\label{2 amplitude}.
\end{align}
\begin{figure*}[t]
\centering
\begin{tabular}{ccc}
\includegraphics[width=0.3\textwidth]{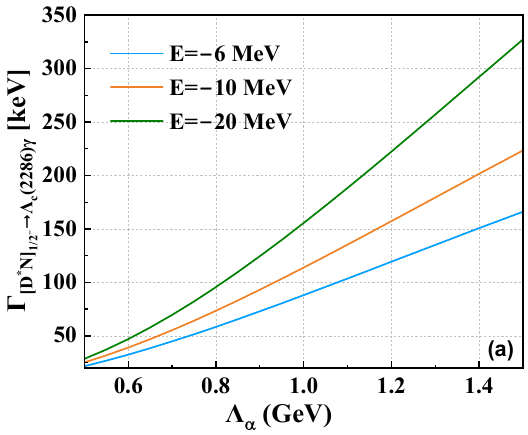}&
\includegraphics[width=0.3\textwidth]{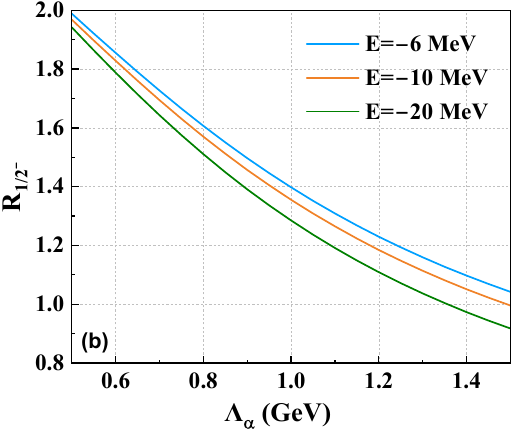}&
\includegraphics[width=0.3\textwidth]{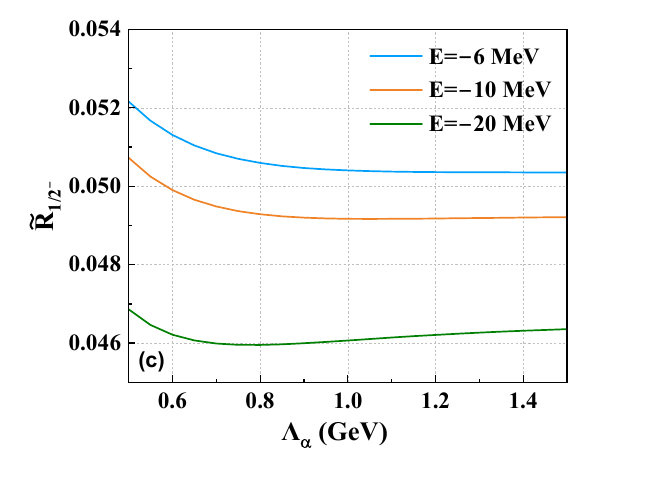}
\end{tabular}
\caption{Radiative decay widths of the $[D^*N]_{1/2^-} \rightarrow \Lambda_c\gamma $ and the ratios $R_{1/2^-}$ and $\tilde{R}_{1/2^-}$ for three typical binding energies $E =-6$ MeV, $-10$ MeV, and $-20$ MeV.}\label{12}
\end{figure*}
Here, $q_i$ represents the four-momentum of the exchanged particle, which can be expressed by
\begin{eqnarray}
q_1&=&p_1-k_1=k_2-p_2,\nonumber\\
q_2&=&p_2-k_1=k_2-p_1,\nonumber\\
q_3&=&p_1+p_2=k_1+k_2.
\end{eqnarray}
While the polarization vectors $\epsilon^\tau_{D^*}$ and $\epsilon^{*\lambda}_{\gamma}$ correspond to the $D^{*}$ charmed meson and photon, respectively. The superscripts ($h$ through $n$) denote the individual diagrams for the process $D^*N \to \Lambda_{c1} \gamma$ shown in Fig. \ref{diagram}. Details regarding the polarization vectors can be found in Ref. \cite{Chen:2024xlw}. Finally, the total scattering amplitudes for the radiative decay processes $D^*N \to \Lambda_c \gamma$, $D^*N \to \Lambda_{c1} \gamma$ and $D^*N \to \Lambda_{c2} \gamma$ are obtained as
\begin{eqnarray}\label{total amplitude}
\hat{\mathcal{M}}^{\rm{total}}_{D^*N\rightarrow \Lambda_{c(2)}\gamma}&=&\sum_{i=a}^{g}{\mathcal{M}^{(i)}},\\
\hat{\mathcal{M}}^{\rm{total}}_{D^*N\rightarrow \Lambda_{c1}\gamma}&=&\sum_{i=h}^{n}{\mathcal{M}^{(i)}}.
\end{eqnarray}
In particular, the total scattering amplitudes for the quantum electrodynamics (QED) process must satisfy gauge invariance when we replace $\epsilon^{*\lambda}_{\gamma}$ with $k_1^\mu$, i.e., $k_1^\mu \mathcal{M}_{\mu}^{\rm total} = 0$. Our calculations ensure that electromagnetic gauge invariance is preserved.

With the above preparations, we can calculate the radiative decay widths for the processes $D^*N \to \Lambda_c \gamma$ and $D^*N \to \Lambda_{c1(2)} \gamma$. The decay amplitude for $D^*N \to \Lambda_c \gamma$ is related to the scattering amplitude as follows \cite{Zhang:2024usz,Luo:2023hnp,Zhang:2006ix}:
\begin{eqnarray}\label{amplitude}
\small{\mathcal{M}^{JM}_{\small{[D^*N]\rightarrow \Lambda_c\gamma}}}&=&\frac{\sqrt{2m_{\small{[D^*N]}}}}{\sqrt{2m_D^*}\sqrt{2m_N}}\nonumber\\
&&\int\frac{\rm d^3\textbf{p}}{(2\pi)^{3/2}}\hat\phi^{JM}_{[D^*N]}(\textbf{p})\otimes\hat{\mathcal{M}}^{\rm total}_{D^*N\rightarrow \Lambda_c\gamma}. 
\end{eqnarray}
 Here, the notation $[D^*N]$ represents a bound state consisting of the charmed meson $D^*$ and the nucleon $N$. As shown in Eq. (\ref{total amplitude}), $\hat{\mathcal{M}}^{\rm total}_{D^*N \rightarrow \Lambda_c\gamma}$ represents total scattering amplitude for the $D^*N \rightarrow \Lambda_c(2286)\gamma$ process, while $\hat{\phi}^{JM}_{[D^*N]}(\textbf{p})$ denotes the momentum-space wave function of the bound state $[D^*N]$. This wave function can be expressed as,
\begin{eqnarray}\label{phiJM}
\hat\phi^{JM}_{[D^*N]}(\textbf{p})&=&\bigg\{\phi_{[D^*N]|^{2S+1}L_J\rangle}(|\textbf{p}|)\,\,C^{S,m_S}_{1m_1,\frac{1}{2}m_2}C^{J,M}_{Sm_S,Lm_L}\,Y_{L,m_L}(\theta,\phi) \nonumber\\
&&\qquad \bigg|\forall ~ L,S,m_L,m_1,m_2\bigg\},
\end{eqnarray}
where $\phi_{[D^*N]|^{2S+1}L_J\rangle}(|\textbf{p}|)$ is the radial wave function in momentum space. As described in Sec. \ref{sec2}, the radial wave function of the $[D^*N]$ molecular state with $J^P = 1/2^-$ and $3/2^-$ is obtained by solving the Schr\"{o}dinger equation. The decay amplitude in Eq. (\ref{amplitude}) should be multiplied by a factor of $1/\sqrt{2}$ to account for the flavor wave function. To incorporate the off-shell effects of the exchanged mesons and suppress contributions from large momenta, a Gaussian form factor $\mathcal{F}(\textbf{p}^2)=\exp(-\textbf{p}^{2n}/\Lambda_\alpha^{2n})$ is introduced where $n=3$. Here,  $\textbf{p}$ denotes the 3-momentum of initial state $D^*$ and the direction of $\textbf{p}$ is specified by $(\theta, \phi)$. $|\textbf{k}|$ represents the 3-momentum of the final state in the center-of-mass frame and it is expressed as $|\textbf{k}| = (m^2_{[D^*N]} - m^2_{\Lambda_c})/2m_{[D^*N]}$. The decay width is then given by
\begin{equation}\label{width}
\Gamma_{[D^*N]\rightarrow  \Lambda_c\gamma}=\frac{1}{2J+1}\frac{|\textbf{k}|}{32\pi^2m_{[D^*N]}^2}\sum_{M}\int \left|\mathcal{M}^{JM}_{[D^*N]\rightarrow \Lambda_c\gamma}\right|^2{\rm d}\Omega_{\textbf{k}},
\end{equation}
where $J$ represents the total angular momentum of the initial bound state. We consider two cases for the total angular momentum: $J = 1/2$ and $J = 3/2$, corresponding to the $[D^*N]$ bound states.
\begin{figure*}[htbp]
\centering
\begin{tabular}{ccc}
\includegraphics[width=0.3\textwidth]{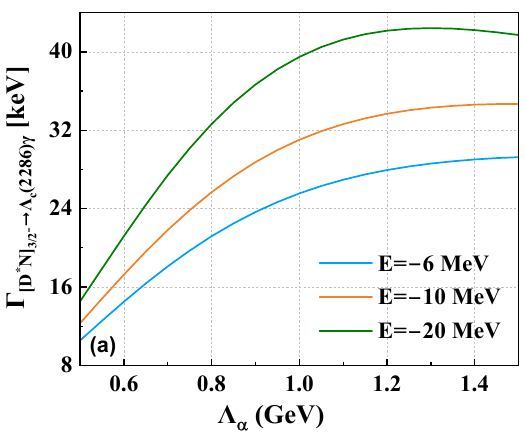}&
\includegraphics[width=0.3\textwidth]{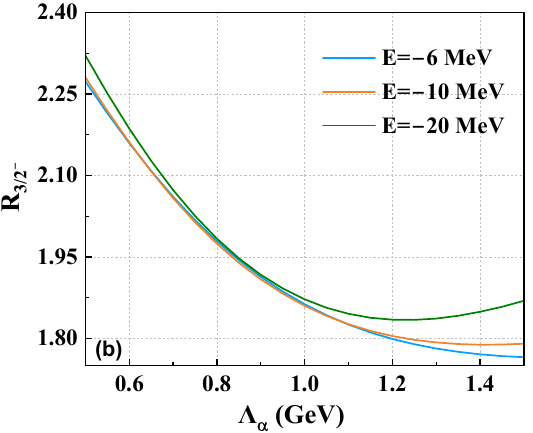}&
\includegraphics[width=0.3\textwidth]{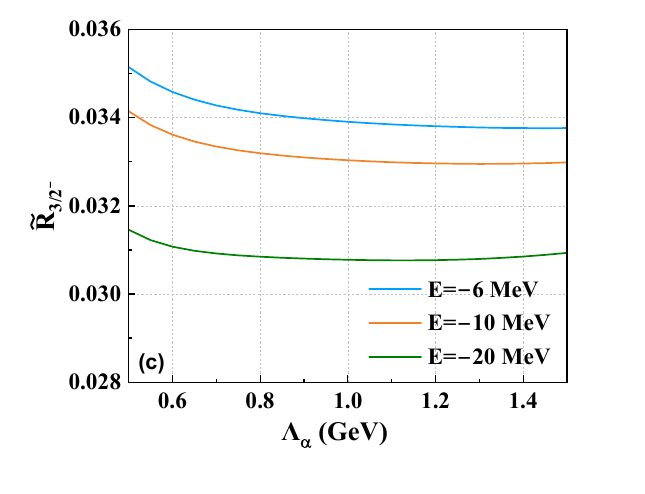}
\end{tabular}
\caption{Radiative decay widths of the $[D^*N]_{3/2^-} \rightarrow \Lambda_c\gamma $ and the ratios $R_{3/2^-}$ and $\tilde{R}_{3/2^-}$ for three typical binding energies $E =-6$ MeV, $-10$ MeV, and $-20$ MeV.}\label{32}
\end{figure*}

We define two ratios involved in the radiative decay widths of $[D^*N]_{1/2^-} \rightarrow \Lambda_{c}(2286)\gamma$, $[D^*N]_{1/2^-} \rightarrow \Lambda_{c}(2595)\gamma$, and $[D^*N]_{1/2^-} \rightarrow \Lambda_{c}(2765)\gamma$, namely,
\begin{eqnarray*}
R_{1/2^-}&=&\frac{\Gamma_{[D^*N]_{1/2^-}\to \Lambda_{c}(2595)\gamma}}{ \Gamma_{[D^*N]_{1/2^-}\to \Lambda_{c}(2286)\gamma}},\\[3pt]
\tilde{R}_{1/2^-}&=&\frac{\Gamma_{[D^*N]_{1/2^-}\to \Lambda_{c}(2765)\gamma}}{ \Gamma_{[D^*N]_{1/2^-}\to \Lambda_{c}(2286)\gamma}}.
\end{eqnarray*}
The conventions of notation $R_{3/2^-}$ and $\tilde{R}_{3/2^-}$ are consistent with $R_{1/2^-}$ and $\tilde{R}_{1/2^-}$ in the following  discussions. In Fig. \ref{12}, we present the radiative decay width of the $[D^*N]_{1/2^-}$ molecular state into $\Lambda_c(2286)\gamma$, along with the ratios $R_{1/2^-}$ and $\tilde{R}_{1/2^-}$ for three typical binding energies: $-6$, $-10$, and $-20$ MeV. These results depend on the cutoff parameter $\Lambda_\alpha$, which is set within the range of 0.5 to 1.5 GeV.

Taking the binding energy of $-6$ MeV as an example, the radiative decay width for the process $[D^*N]_{1/2^-} \rightarrow \Lambda_c(2286)\gamma$ lies within the range of 21.6 to 166.1 keV, while the ratio $R_{1/2^-}$ is approximately 1 to 2. This result indicates that the radiative decay width for the $[D^*N]_{1/2^-} \rightarrow \Lambda_c(2286)\gamma$ process is slightly smaller than that of the $[D^*N]_{1/2^-} \rightarrow \Lambda_c(2595)\gamma$ process. Notably, this conclusion almost remains unchanged even as the absolute value of the binding energy increases. We now examine the radiative decay of the process $[D^*N]_{1/2^-} \to \Lambda_c(2765)\gamma$, the radiative decay widths vary from 1.1 to 8.4 keV and the ratio $\tilde{R}_{1/2}$ changes from 0.050 to 0.052. The smaller radiative decay width of the $[D^*N]_{1/2^-} \to \Lambda_c(2765)\gamma$ process relative to the $[D^*N]_{1/2^-} \to \Lambda_c(2286)\gamma$ process primarily stems from phase space constraints and a relatively weaker coupling strength. As the absolute value of the binding energy increases, the ratio $\tilde{R}_{1/2^-}$ tends to decrease as illustrated in the right plot of Fig. \ref{12}.

A similar analysis is presented in Fig. \ref{32} for the $[D^*N]_{3/2^-}$ molecular state. When binding energy is $-6$ MeV for the $[D^*N]_{3/2^-}$ bound state, the radiative decay width of $[D^*N]_{3/2^-}$ to $\Lambda_{c}(2286)\gamma$ is in the range of 10.5$-$29.2 keV and the corresponding ratio $R_{3/2^-}$ falls between 1.8 and 2.3. Our calculation shows that the radiative decay widths for $[D^*N]_{3/2^-}\to\Lambda_c(2595)\gamma$
and $[D^*N]_{3/2^-}\to\Lambda_c(2286)\gamma$
lie within the same order of magnitude. The ratio $\tilde{R}_{3/2}$ is in the range of 0.033$-$0.035, and the $\Lambda_{c}(2765)\gamma$ channel is suppressed for the $[D^*N]_{3/2^-}$ bound state.

Among the Feynman diagrams presented in Fig. \ref{diagram}, we find that the contributions from Feynman diagrams $(a)$ and $(f)$ are dominant in the decay width of the process $[D^*N] \to \Lambda_c(2286)\gamma$, while the contributions from contact terms are comparatively small. For the processes of $[D^*N]\to\Lambda_c(2595)\gamma$, the Feynman diagram with subscript $(i)$ plays a significant role compared to the process of $[D^*N]\to\Lambda_c(2286)\gamma$. The reason is that parity conservation requires the relative orbital angular momentum between meson $D$ and nucleon $N$ to be in an $S$-wave for vertex $DN\Lambda_{c(1)}$ in the process of $[D^*N]\to\Lambda_c(2595)\gamma$ but a $P$-wave for $[D^*N]\to\Lambda_c(2286)\gamma$. So the diagram with subscript $(b)$ is suppressed naturally for the $[D^*N]\to\Lambda_c(2286)\gamma$ process.

The determination of the couplings $g_{D^{(*)}N\Lambda_{c(1,2)}}$ in Table \ref{coupling} is a little model dependent, which brings some uncertainties to the radiative decay widths. We further examined the dependence of our results on the coupling constants. If we use the coupling constants $g_{DN\Lambda_c}=13.5$ and $g_{D^*N\Lambda_c}=-5.6$ determined by SU(4) symmetry \cite{Lin:1999ve,Liu:2001ce,Dong:2009tg}, the radiative decay widths $\Gamma_{[D^*N]_{1/2^-}\to \Lambda_c\gamma}$ and $\Gamma_{[D^*N]_{3/2^-}\to \Lambda_c\gamma}$ are approximately 25$-$196 keV and 12.6$-$34.3 keV with the binding energy $-6$ MeV, respectively. The corresponding ratios $R_{1/2^-}$ and $\tilde{R}_{1/2^-}$ fall within the ranges of 0.9$-$1.7 and 0.043$-$0.045, while the ratios $R_{3/2^-}$ and $\tilde{R}_{3/2^-}$ are predicted to lie within the ranges of 1.5$-$1.9 and 0.028$-$0.029, correspondingly. The results are very close to ours. When $g_{DN\Lambda_c}=7.9$ and $g_{D^*N\Lambda_c}=-7.5$ are taken from the QCD sum rule \cite{Khodjamirian:2011jp,Navarra:1998vi,Duraes:2000js},
the radiative decay width $\Gamma_{[D^*N]_{1/2^-}\to \Lambda_c\gamma}$ are about 29.8$-$229.2 keV and the $\Gamma_{[D^*N]_{3/2^-}\to \Lambda_c\gamma}$ changes to 14.5$-$40.8 keV. We can find that the radiative decay widths exhibit a weak dependence on the coupling constants $g_{DN\Lambda_c}$ and $g_{D^*N\Lambda_c}$ using different theoretical approaches \cite{Lin:1999ve,Liu:2001ce,Dong:2009tg,Khodjamirian:2011jp,Navarra:1998vi,Duraes:2000js}.

As illustrated in the left plots of Figs. \ref{12} and \ref{32}, the decay width of $[D^*N]_{1/2^-} \to \Lambda_c(2286)\gamma$ is larger than that of $[D^*N]_{3/2^-} \to \Lambda_c(2286)\gamma$. The same trend is observed in the $\Lambda_c(2595)\gamma$ and $\Lambda_c(2765)\gamma$ decay channels. This considerable difference is caused by the requirement that the coupled quantum numbers of the $D^*$ meson and nucleon $N$ should align with the initial state of $\Lambda_c(2940)$, which has $J^P = 1/2^-$ or $J^P = 3/2^-$, as shown in Eq.~(\ref{phiJM}). The significant differences in these two decay processes provide a possibility for identifying the $J^P$ quantum numbers of $\Lambda_c(2940)$.

In addition, the $\Lambda_c(2940)$ can be interpreted as the 2$P$-wave singly charmed baryon $\Lambda_c(2P,1/2^-)$ or $\Lambda_c(2P,3/2^-)$ in quenched picture \cite{Yang:2023fsc,Chen:2014nyo,Lu:2018utx,Ebert:2011kk,Capstick:1986ter,Garcilazo:2007eh,Kim:2021ywp,Zhang:2024afw,Gong:2021jkb}. The coupling of the photon to the constituent $D^*$ meson and $N$ nucleon of the $\Lambda_c(2940)$ is essentially different from that of the quark models, in which the photon directly couples to the quark system\cite{Koniuk:1979vy}. We have calculated the radiative decay width of $\Lambda_c(2940)$ decay to $\Lambda_c(2286)\gamma$, $\Lambda_c(2595)\gamma$, and $\Lambda_c(2765)\gamma$ by regarding $\Lambda_c(2940)$ as a $\Lambda_c(2P,1/2^-)$ or $\Lambda_c(2P,3/2^-)$, employing the approach developed in Refs. \cite{Peng:2024pyl,Deng:2016stx}.
The results are shown in Table \ref{bare}. Similar ratios to those defined earlier can be introduced for these decay channels, which can be defined below,
\begin{eqnarray*}
 R^b_{1/2}&=&\frac{\Gamma_{\Lambda_c(2P,1/2^-)\to \Lambda_{c}(2595)\gamma}}{ \Gamma_{\Lambda_c(2P,1/2^-)\to \Lambda_{c}(2286)\gamma}},\\[3pt]
\tilde{R}^b_{1/2}&=&\frac{\Gamma_{\Lambda_c(2P,1/2^-)\to \Lambda_{c}(2765)\gamma}}{ \Gamma_{\Lambda_c(2P,1/2^-)\to \Lambda_{c}(2286)\gamma}}.
\end{eqnarray*}
The notations $R^b_{3/2}$ and $\tilde{R}^b_{3/2}$ follow the same convention as $R^b_{1/2}$ and $\tilde{R}^b_{1/2}$ in the following discussions.

\vspace{0.1cm}
The results show that the radiative decay widths of $\Lambda_c(2940)$ in the conventional baryon picture are suppressed by at least one to two orders of magnitude compared to those predicted within the $D^*N$ molecular configuration for identical decay channels. In the framework of molecular state, the radiative decay widths of the processes $[D^*N]_{1/2^-} \rightarrow \Lambda_c(2286)\gamma$ and $[D^*N]_{1/2^-} \rightarrow \Lambda_c(2595)\gamma$ are comparable and the ratio $R_{1/2}$ ranges from 1 to 2. However, the situation has changed in a conventional three-quark picture. When $\Lambda_c(2940)$ is regarded as a $\Lambda_c(2P,1/2^-)$ charmed baryon, the ratio $R^b_{1/2}$ is 0.0016 and the process $\Lambda_c(2P,1/2^-) \to \Lambda_c(2595)\gamma$ is suppressed significantly. For the decay $\Lambda_c(2P,1/2^-)\to \Lambda_c(2765)\gamma$, the ratio $\tilde{R}_{1/2}=0.05$ is smaller than $\tilde{R}^b_{1/2}=0.085$. These discrepancies likely stem from distinct photon coupling dynamics between constituent quarks and hadronic molecule systems.

\renewcommand{\arraystretch}{1.6}
\setlength{\tabcolsep}{1.0mm}
\begin{table}[h]
	\centering
        \small
	\caption{Radiative decay width of $\Lambda_c(2940)$ as a conventional charmed baryon.}\label{bare}
	\begin{tabular}{cccc}
		\toprule[1pt]
		Processes & Width & Processes & Width\\
		\midrule[0.75pt]
		$\Lambda_c(2P,1/2^-)\to\Lambda_c\gamma$&1.89 keV&
		$\Lambda_c(2P,3/2^-)\to\Lambda_c\gamma$&2.27 keV\\ $\Lambda_c(2P,1/2^-)\to\Lambda_{c1}\gamma$&3.1eV&
		$\Lambda_c(2P,3/2^-)\to\Lambda_{c1}\gamma$&5.29 keV\\
		$\Lambda_c(2P,1/2^-)\to\Lambda_{c2}\gamma$&0.16 keV&
		$\Lambda_c(2P,3/2^-)\to\Lambda_{c2}\gamma$&0.35 keV\\
		\bottomrule[1pt]
		\bottomrule[1pt]
	\end{tabular}
\end{table}

For the $\Lambda_c(2940)$ with $J^P=3/2^-$, the radiative decay width of $\Lambda_c(2P,3/2^-) \to \Lambda_{c}(2595)\gamma$ exceeds those of the other two decay channels. The ratios $R^b_{3/2}$ and $\tilde{R}^b_{3/2}$ are 2.33 and 0.15, respectively. These values differ from the corresponding ratios for the $D^*N$ molecular states with the same $J^P$. Notably, the ratios $R^b_{1/2}$ and $\tilde{R}^b_{3/2}$ demonstrate striking discrepancies compared to $R_{1/2}$ and $\tilde{R}_{3/2}$, respectively. These comparisons highlight that electromagnetic probes can be effectively utilized to explore the internal structure of hadrons.

\vspace{0.1cm}
The above findings suggest that analyzing the radiative decay behavior of $S$-wave $D^*N$ molecular states can serve as an effective method for determining their $J^P$ quantum numbers, especially the $\Lambda_{c}(2286)\gamma$ and $\Lambda_{c}(2595)\gamma$ channel. The radiative decays of the $\Lambda_c(2940)$ provide significant insights into its internal structure. However, precise determination of the couplings remains challenging due to sparse experimental information and model dependence, which may lead to possible uncertainties in the predictions of radiative decay in this work. In order to eliminate the longstanding debate surrounding the $\Lambda_c(2940)$ and clarify its nature, we expect that the experimental and theoretical colleagues can be inspired from the present work and further decode its properties through more precise measurements and advanced theoretical models.

\section{Summary}\label{sec5}
In this work, we propose that investigating the radiative decay modes $\Lambda_c(2286)\gamma$, $\Lambda_c(2595)\gamma$, and $\Lambda_c(2765)\gamma$ of the newly discovered hadronic state $\Lambda_c(2940)$—particularly the ratios of their decay widths—can provide insights into the internal structure of the $\Lambda_c(2940)$. This approach is motivated by the similarities between the $\Lambda_c(2940)$ and the $X(3872)$, as both states exhibit a low-mass puzzle \cite{Godfrey:1985xj,Yang:2023fsc,Chen:2014nyo,Lu:2018utx,Ebert:2011kk,Capstick:1986ter,Garcilazo:2007eh,Kim:2021ywp}. Moreover, experimental findings highlight the importance of radiative decays of the $X(3872)$ into $J/\psi$ and $\psi(3686)$ for probing its properties \cite{BaBar:2008flx,LHCb:2014jvf,Belle:2011wdj,BESIII:2020nbj,LHCb:2024tpv}. A detailed calculation reveals significant discrepancies in the ratios of the radiative decay widths for the $\Lambda_c(2940)$, depending on the structural assumptions made about the state.

We believe that the insights gained from this work could encourage our experimental colleagues to further investigate the radiative decays of the $\Lambda_c(2940)$. With its upcoming high-luminosity upgrade, LHCb, along with the ongoing Belle II experiment, holds significant potential to probe these decay channels.

In fact, this work further emphasizes the importance of electromagnetic probes, as they are crucial for revealing the internal structure of hadronic states and should not be overlooked in comparison to probes involving strong interactions. As we enter the era of higher precision in the study of hadron spectroscopy, we believe that a multifaceted approach, combining different perspectives, will be essential for advancing our understanding—not only of these novel phenomena but also of the nonperturbative nature of strong interaction.

\section*{ACKNOWLEDGMENTS}
The authors express sincere gratitude to Prof. Zhan-Wei Liu and Prof. Xiang Liu for their insightful discussions and extend thanks to Si-Qiang Luo and Fu-Lai Wang for their valuable contributions to the discussions. This work is supported by the National Natural Science Foundation of China under Grants No. 12175091, No. 12335001, and No. 12247101,  the ‘111 Center’ under Grant No. B20063, the Natural Science Foundation of Gansu Province ( Grants No. 22JR5RA389 and No. 25JRRA799), the innovation project for young science and technology talents of Lanzhou city under Grant No. 2023-QN-107, the fundamental Research Funds for the Central Universities (Grant No. lzujbky-2023-stlt01), and the project for top-notch innovative talents of Gansu province.

\section*{DATA AVAILABILITY}
 The data that support the findings of this article are
 openly available \cite{data}, embargo periods may apply.

\section*{APPENDIX: DETERMINATIONS\\ OF THE COUPLING CONSTANTS}
The coupling constants are determined by the quark-pair-creation model. Generally, the transition matrix of the process of $\Lambda_{c(1,2)}\to D^{(*)}N$ can be expressed as
\begin{eqnarray}
\mathcal{M}_{\rm QPC}(\textbf{p})=\langle D^{(*)}N|\,\mathcal{T}\,|\Lambda_{c(1,2)}\rangle.
\end{eqnarray}
Here, the transition operator $\mathcal{T}$ is
\begin{eqnarray}
\mathcal{T}=-3\gamma\sum_{m}\langle 1 m,1 -m,0 0 \rangle \int d^3\textbf{p}_4 d^3\textbf{p}_5\delta(\textbf{p}_4+\textbf{p}_5) \nonumber \\
\times\mathcal{Y}_1^m(\frac{\textbf{p}_4-\textbf{p}_5}{2})\chi_{1,-m}^{45}\omega_0^{45}\phi_0^{45}b^\dagger_4(\textbf{p}_4)d^\dagger_5(\textbf{p}_5),
\end{eqnarray}
where the dimensionless parameter $\gamma$ describes the strength of the
quark and antiquark pair creation from the vacuum, it can be determined as 9.58 by fitting the total decay width of $\Sigma_c(2520)$ \cite{ParticleDataGroup:2024cfk}.  $\mathcal{Y}_1^m(\textbf{p})=|\textbf{p}|Y_{1,m}(\theta,\phi)$ is a solid harmonic polynomial. $\chi_0^{45}$, $\phi_0^{45}=(u\bar{u}+d\bar{d}+s\bar{s})/\sqrt{3}$, and $\omega_0^{45}=\delta_{ij}$ are spin, flavor, and color wave functions, respectively. $b^\dagger_4$ and $d^\dagger_5$ are quark and antiquark creation operators.

The transition matrix can be further expressed by
\begin{align}
&\langle D^{(*)}N|\,\mathcal{T}\,|\Lambda_{c(1,2)}\rangle(A\to BC) \nonumber\\
&=\gamma\sum_{\substack{M_{L_A},M_{S_A},\\M_{L_B},M_{S_B},\\M_{L_C},M_{S_C},m}}\langle L_{A} M_{L_A} S_A M_{S_A}|J_A M_{J_A}\rangle  \nonumber\\
&\times\langle L_B M_{L_B} S_B M_{S_B}|J_B M_{JB}\rangle \langle L_C M_{L_C} S_C M_{S_C}|J_C M_{J_C}\rangle  \nonumber\\
&\times\langle1 m;1,-m|00\rangle\langle\chi_{S_B M_{SB}}\chi_{S_C M_{S_C}}\chi_{S_A M_{S_A}}|\chi_{1-m}\rangle  \nonumber\\
&\times\langle\phi_B\phi_C|\phi_A\phi_0\rangle I^{M_{L_A,m}}_{M_{L_B}M_{L_C}}(\textbf{p}),
\end{align}
where $I^{M_{L_A,m}}_{M_{L_B}M_{L_C}}(\textbf{p})$ denotes the overlap integral of the wave functions in the momentum space and $\textbf{p}$ is the momentum of an outgoing particle, while the notations $A$, $B$, and $C$ refer to the $\Lambda_{c(1,2)}$, $D^{(*)}$, and $N$, respectively. We use harmonic oscillator spatial wave functions to describe both baryons and mesons. The masses of light and heavy quarks are $m_n=0.37$ GeV and $m_c=1.88$ GeV, respectively. In the center-of-mass system of singly charmed baryon $\Lambda_{c(1,2)}$, the three-momentum of $D^{(*)}$ and $N$ are equal in magnitude and opposite in direction.

We derive the coupling constants $g_{D^{(*)}N\Lambda_{c(1,2)}}$ by comparing the decay widths from quark and hadron levels. We reduce the mass of the nucleon to make the phase space allowed for the $\Lambda_{c(1,2)}\to D^{(*)}N$ process.  We take $\textbf{p}\to 0$ to obtain the couplings in Table \ref{coupling}. The relative signs of the couplings are determined by comparing the amplitudes at the quark and hadron levels.

\vfil



\begin{thebibliography}{200}

\bibitem{Belle:2003nnu}
S.~K.~Choi \textit{et al.} (Belle Collaboration),
Observation of a narrow charmonium-like state in exclusive $B^\pm \to K^\pm \pi^+ \pi^- J/\psi$ decays,
\href{https://journals.aps.org/prl/abstract/10.1103/PhysRevLett.91.262001}{Phys. Rev. Lett. \textbf{91}, 262001 (2003)}.

\bibitem{CDF:2003cab}
D.~Acosta \textit{et al.} (CDF Collaboration),
Observation of the narrow state $X(3872) \to J/\psi \pi^+ \pi^-$ in $\bar{p}p$ collisions at $\sqrt{s} = 1.96$ TeV,
\href{https://doi.org/10.1103/PhysRevLett.93.072001}{Phys. Rev. Lett. \textbf{93}, 072001 (2004)}.

\bibitem{D0:2004zmu}
V.~M.~Abazov \textit{et al.} (D0 Collaboration),
Observation and properties of the $X(3872)$ decaying to $J/\psi \pi^+ \pi^-$ in $p\bar{p}$ collisions at $\sqrt{s} = 1.96$ TeV,
\href{https://doi.org/10.1103/PhysRevLett.93.162002}{Phys. Rev. Lett. \textbf{93}, 162002 (2004)}.

\bibitem{BaBar:2004iez}
B.~Aubert \textit{et al.} (BABAR Collaboration),
Observation of the decay $B \to J/\psi \eta K$ and search for $X(3872) \to J/\psi \eta$,
\href{https://doi.org/10.1103/PhysRevLett.93.041801}{Phys. Rev. Lett. \textbf{93}, 041801 (2004)}.

\bibitem{Swanson:2006st}
E.~S.~Swanson,
The new heavy mesons: A status report,
\href{https://doi.org/10.1016/j.physrep.2006.04.003}{Phys. Rep. \textbf{429}, 243 (2006)}.

\bibitem{Chen:2016qju}
H.~X.~Chen, W.~Chen, X.~Liu, and S.~L.~Zhu,
The hidden-charm pentaquark and tetraquark states,
\href{https://doi.org/10.1016/j.physrep.2016.05.004}{Phys. Rep. \textbf{639}, 1 (2016)}.

\bibitem{Chen:2016spr}
H.~X.~Chen, W.~Chen, X.~Liu, Y.~R.~Liu and S.~L.~Zhu,
A review of the open charm and open bottom systems,
\href{https://iopscience.iop.org/article/10.1088/1361-6633/aa6420}{Rep. Prog. Phys. \textbf{80}, 076201 (2017)}.

\bibitem{Esposito:2016noz}
A.~Esposito, A.~Pilloni and A.~D.~Polosa,
Multiquark resonances,
\href{https://doi.org/10.1016/j.physrep.2016.11.002}{Phys. Rep. \textbf{668}, 1 (2017)}.

\bibitem{Olsen:2017bmm}
S.~L.~Olsen, T.~Skwarnicki and D.~Zieminska,
Nonstandard heavy mesons and baryons: Experimental evidence,
\href{https://journals.aps.org/rmp/pdf/10.1103/RevModPhys.90.015003}{Rev. Mod. Phys. \textbf{90}, 015003 (2018)}.


\bibitem{Brambilla:2019esw}
N.~Brambilla, S.~Eidelman, C.~Hanhart, A.~Nefediev, C.~P.~Shen, C.~E.~Thomas, A.~Vairo, and C.~Z.~Yuan,
The $XYZ$ states: Experimental and theoretical status and perspectives,
\href{https://doi.org/10.1016/j.physrep.2020.05.001}{Phys. Rep. \textbf{873}, 1 (2020)}.


\bibitem{Liu:2019zoy}
Y.~R.~Liu, H.~X.~Chen, W.~Chen, X.~Liu and S.~L.~Zhu,
Pentaquark and Tetraquark states,
\href{https://doi.org/10.1016/j.ppnp.2019.04.003}{Prog. Part. Nucl. Phys. \textbf{107}, 237 (2019)}.


\bibitem{Chen:2022asf}
H.~X.~Chen, W.~Chen, X.~Liu, Y.~R.~Liu and S.~L.~Zhu,
An updated review of the new hadron states,
\href{https://iopscience.iop.org/article/10.1088/1361-6633/aca3b6}{Rep. Prog. Phys. \textbf{86}, 026201 (2023)}.


\bibitem{Liu:2024uxn}
M.~Z.~Liu, Y.~W.~Pan, Z.~W.~Liu, T.~W.~Wu, J.~X.~Lu and L.~S.~Geng,
Three ways to decipher the nature of exotic hadrons: Multiplets, three-body hadronic molecules, and correlation functions,
\href{https://doi.org/10.1016/j.physrep.2024.12.001}{Phys. Rep. \textbf{1108}, 1 (2025)}.


\bibitem{BaBar:2006itc}
B.~Aubert \textit{et al.}. (BABAR Collaboration),
Observation of a charmed baryon decaying to $D^0p$ at a mass near 2.94 $\rm{GeV/c^2}$,
\href{https://doi.org/10.1103/PhysRevLett.98.012001}{Phys. Rev. Lett. \textbf{98}, 012001 (2007)}.


\bibitem{Belle:2006xni}
K.~Abe \textit{et al.} (Belle Collaboration),
Experimental constraints on the possible $J^P$ quantum numbers of the $\Lambda_c(2880)^+$,
\href{https://doi.org/10.1103/PhysRevLett.98.262001}{Phys. Rev. Lett. \textbf{98}, 262001 (2007)}.


\bibitem{Yang:2023fsc}
H.~M.~Yang and H.~X.~Chen,
2$P$-wave charmed baryons from QCD sum rules,
\href{https://doi.org/10.1103/PhysRevD.109.036032}{Phys. Rev. D \textbf{109}, 036032 (2024)}.


\bibitem{Chen:2014nyo}
B.~Chen, K.~W.~Wei and A.~Zhang,
Assignments of $\Lambda_Q$ and $\Xi_Q$ baryons in the heavy quark-light diquark picture,
\href{https://doi.org/10.1140/epja/i2015-15082-3}{Eur. Phys. J. A \textbf{51}, 82 (2015)}.


\bibitem{Lu:2018utx}
Q.~F.~L\"u, L.~Y.~Xiao, Z.~Y.~Wang and X.~H.~Zhong,
Strong decay of $\Lambda _c(2940)$ as a 2$P$ state in the $\Lambda _c$ family,
\href{https://doi.org/10.1140/epjc/s10052-018-6083-7}{Eur. Phys. J. C \textbf{78}, 599 (2018)}.


\bibitem{Ebert:2011kk}
D.~Ebert, R.~N.~Faustov, and V.~O.~Galkin,
Spectroscopy and Regge trajectories of heavy baryons in the relativistic quark-diquark picture,
\href{https://doi.org/10.1103/PhysRevD.84.014025}{Phys. Rev. D \textbf{84}, 014025 (2011)}.


\bibitem{Capstick:1986ter}
S.~Capstick and N.~Isgur,
Baryons in a relativized quark model with chromodynamics,
\href{https://doi.org/10.1103/physrevd.34.2809}{Phys. Rev. D \textbf{34}, 2809 (1986)}.


\bibitem{Garcilazo:2007eh}
H.~Garcilazo, J.~Vijande, and A.~Valcarce,
Faddeev study of heavy baryon spectroscopy,
\href{https://doi.org/10.1088/0954-3899/34/5/014}{J. Phys. G \textbf{34}, 961 (2007)}.


\bibitem{Kim:2021ywp}
Y.~Kim, Y.~R.~Liu, M.~Oka, and K.~Suzuki,
Heavy baryon spectrum with chiral multiplets of scalar and vector diquarks,
\href{https://doi.org/10.1103/PhysRevD.104.054012}{Phys. Rev. D \textbf{104}, 054012 (2021)}.


\bibitem{Godfrey:1985xj}
S.~Godfrey and N.~Isgur,
Mesons in a relativized quark model with chromodynamics,
\href{https://doi.org/10.1103/PhysRevD.32.189}{Phys. Rev. D \textbf{32}, 189 (1985)}.


\bibitem{Swanson:2003tb}
E.~S.~Swanson,
Short range structure in the $X(3872)$,
\href{https://doi.org/10.1016/j.physletb.2004.03.033}{Phys. Lett. B \textbf{588}, 189 (2004)}.

\bibitem{Wong:2003xk}
C.~Y.~Wong,
Molecular states of heavy quark mesons,
\href{https://journals.aps.org/prc/abstract/10.1103/PhysRevC.69.055202}{Phys. Rev. C \textbf{69}, 055202 (2004)}.


\bibitem{Close:2003sg}
F.~E.~Close and P.~R.~Page,
The $D^{*0}$ anti-$D^0$ threshold resonance,
\href{https://doi.org/10.1016/j.physletb.2003.10.032}{Phys. Lett. B \textbf{578}, 119 (2004)}.


\bibitem{Voloshin:2003nt}
M.~B.~Voloshin,
Interference and binding effects in decays of possible molecular component of $X(3872)$,
\href{https://doi.org/10.1016/j.physletb.2003.11.014}{Phys. Lett. B \textbf{579}, 316 (2004)}.


\bibitem{Liu:2008fh}
Y.~R.~Liu, X.~Liu, W.~Z.~Deng, and S.~L.~Zhu,
Is $X(3872)$ really a molecular state?,
\href{https://link.springer.com/article/10.1140/epjc/s10052-008-0640-4}{Eur. Phys. J. C \textbf{56}, 63 (2008)}.


\bibitem{Liu:2009qhy}
X.~Liu, Z.~G.~Luo, Y.~R.~Liu, and S.~L.~Zhu,
$X(3872)$ and other possible heavy molecular states,
\href{https://link.springer.com/article/10.1140/epjc/s10052-009-1020-4}{Eur. Phys. J. C \textbf{61}, 411 (2009)}.


\bibitem{AlFiky:2005jd}
M.~T.~AlFiky, F.~Gabbiani, and A.~A.~Petrov,
$X(3872)$: Hadronic molecules in effective field theory,
\href{https://doi.org/10.1016/j.physletb.2006.07.069}{Phys. Lett. B \textbf{640}, 238 (2006)}.


\bibitem{Thomas:2008ja}
C.~E.~Thomas and F.~E.~Close,
Is $X(3872)$ a molecule?,
\href{https://journals.aps.org/prd/abstract/10.1103/PhysRevD.78.034007}{Phys. Rev. D \textbf{78}, 034007 (2008)}.


\bibitem{Tornqvist:2004qy}
N.~A.~Tornqvist,
Isospin breaking of the narrow charmonium state of Belle at 3872-MeV as a deuson,
\href{https://doi.org/10.1016/j.physletb.2004.03.077}{Phys. Lett. B \textbf{590}, 209 (2004)}.


\bibitem{Lee:2009hy}
I.~W.~Lee, A.~Faessler, T.~Gutsche, and V.~E.~Lyubovitskij,
$X(3872)$ as a molecular $D\bar{D}^*$ state in a potential model,
\href{https://journals.aps.org/prd/abstract/10.1103/PhysRevD.80.094005}{Phys. Rev. D \textbf{80}, 094005 (2009)}.


\bibitem{Braaten:2010mg}
E.~Braaten, H.~W.~Hammer, and T.~Mehen,
Scattering of an ultrasoft Pion and the $X(3872)$,
\href{https://journals.aps.org/prd/abstract/10.1103/PhysRevD.82.034018}{Phys. Rev. D \textbf{82}, 034018 (2010)}.


\bibitem{Wang:2013kva}
P.~Wang and X.~G.~Wang,
Study on $X(3872)$ from effective field theory with pion exchange interaction,
\href{https://journals.aps.org/prl/abstract/10.1103/PhysRevLett.111.042002}{Phys. Rev. Lett. \textbf{111}, 042002 (2013)}.


\bibitem{Baru:2013rta}
V.~Baru, E.~Epelbaum, A.~A.~Filin, C.~Hanhart, U.~G.~Meissner, and A.~V.~Nefediev,
Quark mass dependence of the $X(3872)$ binding energy,
\href{https://doi.org/10.1016/j.physletb.2013.08.073}{Phys. Lett. B \textbf{726}, 537 (2013)}.


\bibitem{Baru:2015nea}
V.~Baru, E.~Epelbaum, A.~A.~Filin, F.~K.~Guo, H.~W.~Hammer, C.~Hanhart, U.~G.~Mei\ss{}ner, and A.~V.~Nefediev,
Remarks on study of $X(3872)$ from effective field theory with pion-exchange interaction,
\href{https://journals.aps.org/prd/abstract/10.1103/PhysRevD.91.034002}{Phys. Rev. D \textbf{91}, 034002 (2015)}.


\bibitem{Song:2023pdq}
J.~Song, L.~R.~Dai, and E.~Oset,
Evolution of compact states to molecular ones with coupled channels: The case of the $X(3872)$,
\href{https://doi.org/10.1103/PhysRevD.108.114017}{Phys. Rev. D \textbf{108}, 114017 (2023)}.


\bibitem{Gamermann:2009uq}
D.~Gamermann, J.~Nieves, E.~Oset, and E.~Ruiz Arriola,
Couplings in coupled channels versus wave functions: Application to the $X(3872)$ resonance,
\href{https://doi.org/10.1103/PhysRevD.81.014029}{Phys. Rev. D \textbf{81}, 014029 (2010)}.


\bibitem{Tornqvist:1993vu}
N.~A.~T\"{o}rnqvist,
On deusons or deuteron-like meson meson bound states,
\href{https://link.springer.com/article/10.1007/BF02734018}{Nuovo Cimento Soc. Ital. Fis. \textbf{107A}, 2471 (1994)}.


\bibitem{He:2006is}
X.~G.~He, X.~Q.~Li, X.~Liu, and X.~Q.~Zeng,
$\Lambda_c(2940)$: A Possible molecular state?,
\href{https://doi.org/10.1140/epjc/s10052-007-0347-y}{Eur. Phys. J. C \textbf{51}, 883 (2007)}.


\bibitem{Dong:2014ksa}
Y.~Dong, A.~Faessler, T.~Gutsche, and V.~E.~Lyubovitskij,
Role of the hadron molecule $\Lambda_c(2940)$ in the $p\bar{p}\to pD^0\bar{\Lambda}_c(2286)$ annihilation reaction,
\href{https://doi.org/10.1103/PhysRevD.90.094001}{Phys. Rev. D \textbf{90}, 094001 (2014)}.


\bibitem{He:2010zq}
J.~He, Y.~T.~Ye, Z.~F.~Sun, and X.~Liu,
The observed charmed hadron $\Lambda_c(2940)$ and the $D^*N$ interaction,
\href{https://doi.org/10.1103/PhysRevD.82.114029}{Phys. Rev. D \textbf{82}, 114029 (2010)}.


\bibitem{Zhang:2012jk}
J.~R.~Zhang,
$S$-wave $D^{(*)}N$ molecular states: $\Sigma_{c}(2800)$ and $\Lambda_{c}(2940)^{+}$?,
\href{https://doi.org/10.1103/PhysRevD.89.096006}{Phys. Rev. D \textbf{89}, 096006 (2014)}.


\bibitem{Yan:2022nxp}
Y.~Yan, X.~Hu, Y.~Wu, H.~Huang, J.~Ping, and Y.~Yang,
Pentaquark interpretation of $\Lambda _{c}$ states in the quark model,
\href{https://doi.org/10.1140/epjc/s10052-023-11709-2}{Eur. Phys. J. C \textbf{83}, 524 (2023)}.


\bibitem{Yan:2023ttx}
M.~J.~Yan, F.~Z.~Peng, and M.~Pavon Valderrama,
Molecular charmed baryons and pentaquarks from light-meson exchange saturation,
\href{https://doi.org/10.1103/PhysRevD.109.014023}{Phys. Rev. D \textbf{109}, 014023 (2024)}.


\bibitem{Xin:2023gkf}
Q.~Xin, X.~S.~Yang, and Z.~G.~Wang,
The singly charmed pentaquark molecular states via the QCD sum rules,
\href{https://doi.org/10.1142/S0217751X23501233}{Int. J. Mod. Phys. A \textbf{38}, 2350123 (2023)}.


\bibitem{Cheng:2006dk}
H.~Y.~Cheng and C.~K.~Chua,
Strong decays of charmed baryons in heavy hadron chiral perturbation theory,
\href{https://doi.org/10.1103/PhysRevD.75.014006}{Phys. Rev. D \textbf{75}, 014006 (2007)}.


\bibitem{Dong:2010xv}
Y.~Dong, A.~Faessler, T.~Gutsche, S.~Kumano and V.~E.~Lyubovitskij,
Radiative decay of $\Lambda_c(2940)$ in a hadronic molecule picture,
\href{https://doi.org/10.1103/PhysRevD.82.034035}{Phys. Rev. D \textbf{82}, 034035 (2010)}.


\bibitem{Wang:2015rda}
X.~Y.~Wang, A.~Guskov and X.~R.~Chen,
$\Lambda_c^*(2940)^+$ photoproduction off the neutron,
\href{https://doi.org/10.1103/PhysRevD.92.094032}{Phys. Rev. D \textbf{92}, 094032 (2015)}.


\bibitem{Dong:2011ys}
Y.~Dong, A.~Faessler, T.~Gutsche, S.~Kumano, and V.~E.~Lyubovitskij,
Strong three-body decays of $\Lambda_c(2940)$,
\href{https://doi.org/10.1103/PhysRevD.83.094005}{Phys. Rev. D \textbf{83}, 094005 (2011)}.


\bibitem{Dong:2009tg}
Y.~Dong, A.~Faessler, T.~Gutsche, and V.~E.~Lyubovitskij,
Strong two-body decays of the $\Lambda_c(2940)$ in a hadronic molecule picture,
\href{https://doi.org/10.1103/PhysRevD.81.014006}{Phys. Rev. D \textbf{81}, 014006 (2010)}.


\bibitem{Ortega:2012cx}
P.~G.~Ortega, D.~R.~Entem, and F.~Fernandez,
Quark model description of the $\Lambda_c(2940)$ as a molecular $D^*N$ state and the possible existence of the$ \Lambda_b(6248)$,
\href{https://doi.org/10.1016/j.physletb.2012.12.025}{Phys. Lett. B \textbf{718}, 1381 (2013)}.


\bibitem{Wang:2020dhf}
B.~Wang, L.~Meng, and S.~L.~Zhu,
$D^{*}N$ interaction and the structure of $\Sigma_c(2800)$ and $\Lambda_c(2940)$ in chiral effective field theory,
\href{https://doi.org/10.1103/PhysRevD.101.094035}{Phys. Rev. D \textbf{101}, 094035 (2020)}.


\bibitem{Yue:2024paz}
Z.~L.~Yue, Q.~Y.~Guo, and D.~Y.~Chen,
Strong decays of the $\Lambda_c(2910)$ and $\Lambda_c(2940)$ in the $D^*N$ molecular frame,
\href{https://doi.org/10.1103/PhysRevD.109.094049}{Phys. Rev. D \textbf{109}, 094049 (2024)}.


\bibitem{Guo:2025efg}
Q.~Y.~Guo and D.~Y.~Chen,
$\Lambda_c(2910)$ and $\Lambda_c(2940)$ productions in $\pi^-p$ scattering process,
\href{https://doi.org/10.1103/pplc-7r3c}{Phys. Rev. D \textbf{111}, 114011 (2025)}.


\bibitem{Ortega:2013fta}
P.~G.~Ortega, D.~R.~Entem, and F.~Fernandez,
The $\Lambda_{c}(2940)^+$ as a $D^*N$ molecule in a constituent quark model and a possible $\Lambda_{b}(6248)$,
\href{https://doi.org/10.1007/s00601-012-0569-x}{Few Body Syst. \textbf{54}, 1101 (2013)}.


\bibitem{Ortega:2014eoa}
P.~G.~Ortega, D.~R.~Entem, and F.~Fern\'andez,
Hadronic molecules in the open charm and open bottom baryon spectrum,
\href{https://doi.org/10.1140/epjc/s10052-025-13891-x}{Phys. Rev. D \textbf{90}, 114013 (2014)}.


\bibitem{Zhang:2019vqe}
D.~Zhang, D.~Yang, X.~F.~Wang, and K.~Nakayama,
Possible $S$-wave $ND^{(*)}$ and $N\bar B^{(*)}$ bound states in a chiral quark model,
\href{https://arxiv.org/abs/1903.01207}{arXiv:1903.01207}.


\bibitem{Belle:2011wdj}
V.~Bhardwaj \textit{et al.} (Belle Collaboration),
Observation of $X(3872)\to J/\psi \gamma$ and search for $X(3872)\to\psi'\gamma$ in B decays,
\href{https://journals.aps.org/prl/abstract/10.1103/PhysRevLett.107.091803}{Phys. Rev. Lett. \textbf{107}, 091803 (2011)}.


\bibitem{BaBar:2008flx}
B.~Aubert \textit{et al.} (\textit{BABAR} Collaboration),
Evidence for $X(3872) \to \psi_{2S} \gamma$ in $B^\pm \to X(3872) K^\pm$ decays, and a study of $B \to c \bar{c} \gamma K$,
\href{https://journals.aps.org/prl/abstract/10.1103/PhysRevLett.102.132001}{Phys. Rev. Lett. \textbf{102}, 132001 (2009)}.


\bibitem{LHCb:2014jvf}
R.~Aaij \textit{et al.} (LHCb Collaboration),
Evidence for the decay $X(3872)\rightarrow\psi(2S)\gamma$,
\href{https://doi.org/10.1016/j.nuclphysb.2014.06.011}{Nucl. Phys. \textbf{B886}, 665 (2014)}.


\bibitem{BESIII:2020nbj}
M.~Ablikim \textit{et al.} (BESIII Collaboration),
Study of open-charm decays and radiative transitions of the $X(3872)$,
\href{https://journals.aps.org/prl/abstract/10.1103/PhysRevLett.124.242001}{Phys. Rev. Lett. \textbf{124}, 242001 (2020)}.


\bibitem{LHCb:2024tpv}
I.~Bezshyiko \textit{et al.} (LHCb Collaboration),
Probing the nature of the $\ensuremath{\chi}_{c1}(3872)$ state using radiative decays,
\href{https://doi.org/10.1007/JHEP11(2024)121}{J.High Energy Phys.11  (2024) 121}.


\bibitem{Barnes:2003vb}
T.~Barnes and S.~Godfrey,
Charmonium options for the $X(3872)$,
\href{https://journals.aps.org/prd/abstract/10.1103/PhysRevD.69.054008}{Phys. Rev. D \textbf{69}, 054008 (2004)}.


\bibitem{Badalian:2012jz}
A.~M.~Badalian, V.~D.~Orlovsky, Y.~A.~Simonov, and B.~L.~G.~Bakker,
The ratio of decay widths of $X(3872)$ to $\psi^{\prime}\gamma $ and $J/\psi\gamma$ as a test of the $X(3872)$ dynamical structure,
\href{https://journals.aps.org/prd/abstract/10.1103/PhysRevD.85.114002}{Phys. Rev. D \textbf{85}, 114002 (2012)}.


\bibitem{Wang:2010ej}
T.~H.~Wang and G.~L.~Wang,
Radiative E1 decays of $X(3872)$,
\href{https://doi.org/10.1016/j.physletb.2011.02.014}{Phys. Lett. B \textbf{697}, 233 (2011)}.


\bibitem{Mehen:2011ds}
T.~Mehen and R.~Springer,
Radiative decays $X(3872) \to \psi(2S)\gamma$ and $\psi(4040)$ $\rightarrow$ $X(3872)\gamma$ in effective field theory,
\href{https://journals.aps.org/prd/abstract/10.1103/PhysRevD.83.094009}{Phys. Rev. D \textbf{83}, 094009 (2011)}.


\bibitem{Li:2009zu}
B.~Q.~Li and K.~T.~Chao,
Higher charmonia and X,Y,Z states with screened potential,
\href{https://journals.aps.org/prd/abstract/10.1103/PhysRevD.79.094004}{Phys. Rev. D \textbf{79}, 094004 (2009)}.


\bibitem{Lahde:2002wj}
T.~A.~Lahde,
Exchange current operators and electromagnetic dipole transitions in heavy quarkonia,
\href{https://doi.org/10.1016/S0375-9474(02)01362-3}{Nucl. Phys.  \textbf{A714}, 183 (2003)}.


\bibitem{Barnes:2005pb}
T.~Barnes, S.~Godfrey, and E.~S.~Swanson,
Higher charmonia,
\href{https://journals.aps.org/prd/abstract/10.1103/PhysRevD.72.054026}{Phys. Rev. D \textbf{72}, 054026 (2005)}.


\bibitem{Yu:2023nxk}
S.~Y.~Yu and X.~W.~Kang,
Nature of $X(3872)$ from its radiative decay,
\href{https://doi.org/10.1016/j.physletb.2023.138404}{Phys. Lett. B \textbf{848}, 138404 (2024)}.


\bibitem{Eichten:2005ga}
E.~J.~Eichten, K.~Lane, and C.~Quigg,
New states above charm threshold,
\href{https://journals.aps.org/prd/abstract/10.1103/PhysRevD.73.014014}{Phys. Rev. D \textbf{73}, 014014 (2006);}
\href{https://journals.aps.org/prd/abstract/10.1103/PhysRevD.73.014014}{\textbf{73}, 079903(E) (2006)]}.


\bibitem{Cardoso:2014xda}
M.~Cardoso, G.~Rupp, and E.~van Beveren,
Unquenched quark-model calculation of $X(3872)$ electromagnetic decays,
\href{https://link.springer.com/article/10.1140/epjc/s10052-014-3254-z}{Eur. Phys. J. C \textbf{75}, 26 (2015)}.


\bibitem{Swanson:2004pp}
E.~S.~Swanson,
Diagnostic decays of the $X(3872)$,
\href{https://doi.org/10.1016/j.physletb.2004.07.059}{Phys. Lett. B \textbf{598}, 197 (2004)}.


\bibitem{Guo:2014taa}
F.~K.~Guo, C.~Hanhart, Y.~S.~Kalashnikova, U.~G.~Mei\ss{}ner, and A.~V.~Nefediev,
What can radiative decays of the $X(3872)$ teach us about its nature?,
\href{https://doi.org/10.1016/j.physletb.2015.02.013}{Phys. Lett. B \textbf{742}, 394 (2015)}.


\bibitem{Dong:2009uf}
Y.~Dong, A.~Faessler, T.~Gutsche, and V.~E.~Lyubovitskij,
$J/\psi\gamma$  and $\psi(2S)\gamma$ decay modes of the $X(3872)$,
\href{https://doi.org/10.1088/0954-3899/38/1/015001}{J. Phys. G \textbf{38}, 015001 (2011)}.


\bibitem{Chen:2024xlw}
P.~Chen, Z.~W.~Liu, Z.~L.~Zhang, S.~Q.~Luo, F.~L.~Wang, J.~Z.~Wang, and X.~Liu,
Role of electromagnetic interactions in the $X(3872)$ and its analogs,
\href{https://doi.org/10.1103/PhysRevD.109.094002}{Phys. Rev. D \textbf{109},094002 (2024)}.


\bibitem{Burdman:1992gh}
G.~Burdman and J.~F.~Donoghue,
Union of chiral and heavy quark symmetries,
\href{https://doi.org/10.1016/0370-2693(92)90068-F}{Phys. Lett. B \textbf{280}, 287 (1992)}.


\bibitem{CLEO:2001foe}
S.~Ahmed \textit{et al.} (CLEO Collaboration),
First measurement of $\Gamma(D^{*+})$,
\href{https://journals.aps.org/prl/abstract/10.1103/PhysRevLett.87.251801}{Phys. Rev. Lett. \textbf{87}, 251801 (2001)}.


\bibitem{Bando:1987br}
M.~Bando, T.~Kugo, and K.~Yamawaki,
Nonlinear realization and hidden local symmetries,
\href{https://doi.org/10.1016/0370-1573(88)90019-1}{Phys. Rep. \textbf{164}, 217 (1988)}.


\bibitem{Isola:2003fh}
C.~Isola, M.~Ladisa, G.~Nardulli, and P.~Santorelli,
Charming penguins in $B\to K^*\pi$, $K(\rho,\omega,\phi)$ decays,
\href{https://doi.org/10.1103/PhysRevD.68.114001}{Phys. Rev. D \textbf{68}, 114001 (2003)}


\bibitem{Casalbuoni:1996pg}
R.~Casalbuoni, A.~Deandrea, N.~Di Bartolomeo, R.~Gatto, F.~Feruglio, and G.~Nardulli,
Phenomenology of heavy meson chiral Lagrangians,
\href{https://doi.org/10.1016/S0370-1573(96)00027-0}{Phys. Rep. \textbf{281}, 145 (1997)}.


\bibitem{Liu:2008xz}
X.~Liu, Y.~R.~Liu, W.~Z.~Deng, and S.~L.~Zhu,
$Z^+(4430)$ as a $D_1^\prime D^* (D_1D^*)$ molecular state,
\href{https://journals.aps.org/prd/abstract/10.1103/PhysRevD.77.094015}{Phys. Rev. D \textbf{77}, 094015 (2008)}.


\bibitem{Tsushima:1998jz}
K.~Tsushima, A.~Sibirtsev, A.~W.~Thomas, and G.~Q.~Li,
Resonance model study of kaon production in baryon baryon reactions for heavy ion collisions,
\href{https://doi.org/10.1103/PhysRevC.59.369}{Phys. Rev. C \textbf{59}, 369 (1999)};
\href{https://doi.org/10.1103/PhysRevC.61.029903}{\textbf{61}, 029903(E) (2000)}.


\bibitem{Engel:1996ic}
A.~Engel, A.~K.~Dutt-Mazumder, R.~Shyam, and U.~Mosel,
Pion production in proton proton collisions in a covariant one boson exchange model,
\href{https://doi.org/10.1016/0375-9474(96)80008-F}{Nucl. Phys. \textbf{A 603}, 387 (1996)}.


\bibitem{Machleidt:2000ge}
R.~Machleidt,
The high precision, charge dependent Bonn nucleon-nucleon potential (CD-Bonn),
\href{https://doi.org/10.1103/PhysRevC.63.024001}{Phys. Rev. C \textbf{63}, 024001 (2001)}.


\bibitem{Cao:2010km}
X.~Cao, B.~S.~Zou, and H.~S.~Xu,
Phenomenological analysis of the double pion production in nucleon-nucleon collisions up to 2.2 GeV,
\href{https://doi.org/10.1103/PhysRevC.81.065201}{Phys. Rev. C \textbf{81}, 065201 (2010)}.


\bibitem{Luo:2022cun}
S.~Q.~Luo, L.~S.~Geng, and X.~Liu,
Double-charm heptaquark states composed of two charmed mesons and one nucleon,
\href{https://doi.org/10.1103/PhysRevD.106.014017}{Phys. Rev. D \textbf{106}, 014017 (2022)}.


\bibitem{ParticleDataGroup:2024cfk}
S.~Navas \textit{et al.} (Particle Data Group),
Review of particle physics,
\href{https://doi.org/10.1103/PhysRevD.110.030001}{Phys. Rev. D \textbf{110}, 030001 (2024)}.


\bibitem{Berestetsky:1982}
V. B. Berestetsky, E. M. Lifshitz, and L. P. Pitaevsky, \textit{Quantum Electrodynamics} (Pergamon Press, New York,
1982).


\bibitem{Tornqvist:1993ng}
N.~A.~T\"{o}rnqvist,
From the deuteron to deusons, an analysis of deuteronlike meson-meson bound states,
\href{https://link.springer.com/article/10.1007/BF01413192}{Z. Phys. C \textbf{61}, 525 (1994)}.


\bibitem{Hiyama:2003cu}
E.~Hiyama, Y.~Kino and M.~Kamimura,
Gaussian expansion method for few-body systems,
\href{https://doi.org/10.1016/S0146-6410(03)90015-9}{Prog. Part. Nucl. Phys. \textbf{51}, 223 (2003)}.


\bibitem{Hiyama:2012}
E.~Hiyama, Y.~Kino and M.~Kamimura,
Gaussian expansion method for few-body systems and its applications to atomic and nuclear physics,
\href{https://doi.org/10.1093/ptep/pts015}{Prog. Theor. Exp. Phys. \textbf{2012}, 01A204 (2012)}.


\bibitem{LHCb:2017jym}
R.~Aaij \textit{et al.} (LHCb Collaboration),
Study of the $D^0 p$ amplitude in $\Lambda_b^0\to D^0 p \pi^-$ decays,
\href{https://doi.org/10.1007/JHEP05(2017)030}{J. High Energy Phys. 05 (2017) 030}.


\bibitem{Chen:2016iyi}
B.~Chen, K.~W.~Wei, X.~Liu, and T.~Matsuki,
Low-lying charmed and charmed-strange baryon states,
\href{https://doi.org/10.1140/epjc/s10052-017-4708-x}{Eur. Phys. J. C \textbf{77},  154 (2017)}.


\bibitem{Wang:2020mxk}
Z.~G.~Wang and H.~J.~Wang,
Analysis of the 1$S$ and 2$S$ states of $\Lambda_Q$ and $\Xi_Q$ with QCD sum rules,
\href{https://doi.org/10.1088/1674-1137/abc1d3}{Chin. Phys. C \textbf{45}, no.1, 013109 (2021)}.


\bibitem{Zhang:2024afw}
Y.~B.~Zhang, L.~Y.~Xiao and X.~H.~Zhong,
Possible explanations of the observed $\Lambda_c$ resonances,
\href{https://doi.org/10.1088/1674-1137/aded0010.1088/1674-1137/aded00}{Chin. Phys. C \textbf{49}, 112001 (2025)}.


\bibitem{Lu:2016ctt}
Q.~F.~L\"u, Y.~Dong, X.~Liu, and T.~Matsuki,
Puzzle of the $\Lambda_c$ spectrum,
\href{https://doi.org/10.11804/NuclPhysRev.35.01.001}{Nucl. Phys. Rev. \textbf{35}, 1 (2018)}.


\bibitem{He:2011jp}
J.~He, Z.~Ouyang, X.~Liu, and X.~Q.~Li,
Production of charmed baryon $\Lambda_c(2940)$ at PANDA,
\href{https://doi.org/10.1103/PhysRevD.84.114010}{Phys. Rev. D \textbf{84}, 114010 (2011)}.


\bibitem{Qiao:2024acm}
K.~S.~Qiao and B.~S.~Zou,
Investigation of $\Lambda_c$ states and $\bar{D}N$ molecule production at the EicC and EIC,
\href{https://doi.org/10.1103/PhysRevD.111.056029}{Phys. Rev. D \textbf{111}, 056029 (2025)}.


\bibitem{Zhang:2022pxc}
Z.~L.~Zhang, Z.~W.~Liu, S.~Q.~Luo, F.~L.~Wang, B.~Wang, and H.~Xu,
$\Lambda_c(2910)^+$ and $\Lambda_c(2940)$ as conventional baryons dressed with the $D^{*}N$ channel,
\href{https://doi.org/10.1103/PhysRevD.107.034036}{Phys. Rev. D \textbf{107}, 034036 (2023)}.


\bibitem{Lin:1999ve}
Z.~w.~Lin, C.~M.~Ko, and B.~Zhang,
Hadronic scattering of charm mesons,
\href{https://doi.org/10.1103/PhysRevC.61.024904}{Phys. Rev. C \textbf{61}, 024904 (2000)}.


\bibitem{Liu:2001ce}
W.~Liu, C.~M.~Ko, and Z.~W.~Lin,
Cross-section for charmonium absorption by nucleons,
\href{https://doi.org/10.1103/PhysRevC.65.015203}{Phys. Rev. C \textbf{65}, 015203 (2002)}.


\bibitem{Khodjamirian:2011jp}
A.~Khodjamirian, C.~Klein, T.~Mannel, and Y.~M.~Wang,
Form factors and strong couplings of heavy baryons from QCD light-cone sum rules,
\href{https://doi.org/10.1007/JHEP09(2011)106}{J. High Energy Phys. 09, (2011) 106}.


\bibitem{Duraes:2000js}
F.~O.~Duraes, F.~S.~Navarra, and M.~Nielsen,
How hard are the form-factors in hadronic vertices with heavy mesons?,
\href{https://doi.org/10.1016/S0370-2693(01)00011-9}{Phys. Lett. B \textbf{498}, 169 (2001)}.


\bibitem{Navarra:1998vi}
F.~S.~Navarra and M.~Nielsen,
$g_{DN\Lambda_c}$ from QCD sum rules,
\href{https://doi.org/10.1016/S0370-2693(98)01247-7}{Phys. Lett. B \textbf{443}, 285 (1998)}.


\bibitem{Haberzettl:2006bn}
H.~Haberzettl, K.~Nakayama, and S.~Krewald,
Gauge-invariant approach to meson photoproduction including the final-state interaction,
\href{https://doi.org/10.1103/PhysRevC.74.045202}{Phys. Rev. C \textbf{74}, 045202 (2006)}.


\bibitem{Huang:2011as}
F.~Huang, M.~Doring, H.~Haberzettl, J.~Haidenbauer, C.~Hanhart, S.~Krewald, U.~G.~Meissner, and K.~Nakayama,
Pion photoproduction in a dynamical coupled-channels model,
\href{https://doi.org/10.1103/PhysRevC.85.054003}{Phys. Rev. C \textbf{85}, 054003 (2012)}.


\bibitem{Gao:2010hy}
P.~Gao, J.~J.~Wu, and B.~S.~Zou,
Possible $\Sigma({1\over2}^-)$ under the $\Sigma^*(1385)$ peak in $K\Sigma^*$ photoproduction,
\href{https://doi.org/10.1103/PhysRevC.81.055203}{Phys. Rev. C \textbf{81}, 055203 (2010)}.


\bibitem{Huang:2016tcr}
Y.~Huang, J.~J.~Xie, J.~He, X.~Chen, and H.~F.~Zhang,
Photoproduction of hidden-charm states in the $\gamma p \to \bar D^{*0} \Lambda^+_c$ reaction near threshold,
\href{https://doi.org/10.1088/1674-1137/40/12/124104}{Chin. Phys. C \textbf{40},  124104 (2016)}.


\bibitem{Luo:2023hnp}
S.~Q.~Luo, Z.~W.~Liu, and X.~Liu,
New type of hydrogenlike charm-pion or charm-kaon matter,
\href{https://doi.org/10.1103/PhysRevD.107.054022}{Phys. Rev. D \textbf{107}, 054022 (2023)}.


\bibitem{Zhang:2006ix}
Y.~J.~Zhang, H.~C.~Chiang, P.~N.~Shen, and B.~S.~Zou,
Possible $S$-wave bound-states of two pseudoscalar mesons,
\href{https://doi.org/10.1103/PhysRevD.74.014013}{Phys. Rev. D \textbf{74}, 014013 (2006)}.


\bibitem{Zhang:2024usz}
Z.~L.~Zhang, Z.~W.~Liu, S.~Q.~Luo, P.~Chen, and Z.~H.~Guo,
Masses and radiative decay widths of $D_{s0}^*(2317)$ and $D_{s1}^\prime(2460)$ and their bottom analogs,
\href{https://doi.org/10.1103/PhysRevD.110.094037}{Phys. Rev. D \textbf{110}, 094037 (2024)}.


\bibitem{Gong:2021jkb}
K.~Gong, H.~Y.~Jing, and A.~Zhang,
Possible assignments of highly excited $\Lambda _c(2860)^+$, $\Lambda _c(2880)^+$ and $\Lambda _c(2940)^+$,
\href{https://doi.org/10.1140/epjc/s10052-021-09255-w}{Eur. Phys. J. C \textbf{81}, 467 (2021)}.

\bibitem{Koniuk:1979vy}
R.~Koniuk and N.~Isgur,
Baryon decays in a quark model with chromodynamics,
\href{https://doi.org/10.1103/PhysRevD.21.1868}{Phys. Rev. D \textbf{21}, 1868 (1980);
\href{https://doi.org/10.1103/PhysRevD.23.818}{\textbf{23}, 818(E) (1981)}}.


\bibitem{Peng:2024pyl}
Y.~X.~Peng, S.~Q.~Luo, and X.~Liu,
Refining radiative decay studies in singly heavy baryons,
\href{https://doi.org/10.1103/PhysRevD.110.074034}{Phys. Rev. D \textbf{110}, 074034 (2024)}.

\bibitem{Deng:2016stx}
W.~J.~Deng, H.~Liu, L.~C.~Gui, and X.~H.~Zhong,
Charmonium spectrum and their electromagnetic transitions with higher multipole contributions,
\href{https://doi.org/10.1103/PhysRevD.95.034026}{Phys. Rev. D \textbf{95},  034026 (2017)}.


\bibitem{data}
P. Chen, Z. L. Zhang, and Y. Zhuge,
The wave functions and radiative decay widths of $\Lambda_c(2940)$,
\href{https://doi.org/10.5281/zenodo.16961102}{10.5281/zenodo.16961102} (2025).

\end{thebibliography}
\end{document}